\documentclass[fleqn,usenatbib]{mnras}

\usepackage{newtxtext,newtxmath}
\usepackage{booktabs}
\usepackage[symbol]{footmisc}
\usepackage[T1]{fontenc}
\usepackage{xspace}

\DeclareRobustCommand{\VAN}[3]{#2}
\let\VANthebibliography\thebibliography
\def\thebibliography{\DeclareRobustCommand{\VAN}[3]{##3}\VANthebibliography}

\usepackage{graphicx}	%
\usepackage{physics,amsmath}	%
\newcommand{\pcm}{cm$^{-2}$}	%
\newcommand{\msun}{\ensuremath{M_{\sun}}}  %
\newcommand{\au}{\ensuremath{\rm AU}}  %
\newcommand{\kms}{\ensuremath{\rm km\,s^{-1}}}  %
\newcommand{\rhounit}{\ensuremath{\msun\,{\rm pc}^{-3}}}  %
\newcommand{\Sigmaunit}{\ensuremath{\msun\,{\rm pc}^{-2}}}  %

\newcommand{\mcl}{\ensuremath{M_{0}}}
\newcommand{\rcl}{\ensuremath{R_{\rm cloud}}}
\newcommand{\alphaturb}{\ensuremath{\alpha_{\rm turb}}}
\newcommand{\alphavir}{\ensuremath{\alpha_{\rm vir}}}

\newcommand{\tclean}{\ensuremath{t_{\rm clean}}}
\newcommand{\texp}{\ensuremath{t_{\rm exp}}}
\newcommand{\tedge}{\ensuremath{t_{\rm init}}}

\newcommand{\vturb}{\ensuremath{\vb{v}_{\rm turb}}}
\newcommand{\mgroup}{\ensuremath{m_{\rm group}}}

\newcommand{\starforge}{\textsc{STARFORGE}\xspace} %
\newcommand{\nbodycode}{\texttt{Nbody6++GPU}\xspace} %
\newcommand{\fiducial}{\texttt{fiducial}\xspace}
\newcommand{\rsmall}{\texttt{R3}\xspace}
\newcommand{\rlarge}{\texttt{R30}\xspace}
\newcommand{\alphaL}{\texttt{alpha1}\xspace}
\newcommand{\alphaH}{\texttt{alpha4}\xspace}
\newcommand{\ISRFten}{\texttt{ISRFx10}\xspace}
\newcommand{\ISRFhundred}{\texttt{ISRFx100}\xspace}
\newcommand{\Bten}{\texttt{Bx10}\xspace}
\newcommand{\Bhundred}{\texttt{Bx100}\xspace}
\newcommand{\Zten}{\texttt{Z01}\xspace}
\newcommand{\Zhundred}{\texttt{Z001}\xspace}

\newcommand{\noextpot}{\texttt{NoExtPot}\xspace}
\newcommand{\extpot}{\texttt{ExtPot}\xspace}

\defcitealias{Grudic2021}{Paper~I}  %
\defcitealias{Guszejnov2021}{Paper~II} %
\defcitealias{Grudic2022}{Paper~III}  %
\defcitealias{Guszejnov2022}{Paper~IV} %
\defcitealias{Guszejnov2022a}{Paper~V} %
\defcitealias{Guszejnov2023}{Paper~VI} %

\title[Stellar Populations in \starforge]{Stellar Populations in STARFORGE: The Origin and
Evolution of Star Clusters and Associations}

\author[Farias et al.]{
Juan P.\ Farias,$^{1}$\thanks{E-mail: juan.farias@austin.utexas.edu}
Stella S.\ R. Offner,$^{1}$
Michael Y. Grudi{\'c}$,^{2,\dagger}$
Dávid Guszejnov,$^{3,\dagger}$
Anna L. Rosen,$^{4, 5, 6}$
\newauthor
and the STARFORGE team \\
$^{1}$Department of Astronomy, University of Texas at Austin, TX 78712 \\
$^{2}$ Carnegie Observatories, 813 Santa Barbara St, Pasadena, CA 91101, USA\\
$^{\dagger}$NASA Hubble Fellow\\
$^{3}$  Center for Astrophysics | Harvard \& Smithsonian, 60 Garden St, Cambridge, MA 02138, USA \\
$^{4}$ Department of Astronomy, San Diego State University, San Diego, CA 92182, USA \\ 
$^{5}$ Computational Science Research Center, San Diego State University, San Diego, CA 92182, USA \\
$^{6}$ Department of Astronomy \& Astrophysics, University of California, San Diego, La Jolla, CA 92093, USA
}

\date{Accepted XXX. Received YYY; in original form ZZZ}

\pubyear{2023}

\begin{document}
\label{firstpage}
\pagerange{\pageref{firstpage}--\pageref{lastpage}}
\maketitle

\begin{abstract}
Most stars form in highly clustered environments within molecular clouds, but
eventually disperse into the distributed stellar field population. Exactly how the
stellar distribution evolves from the embedded stage into gas-free associations and
(bound) clusters is poorly understood. We investigate the long-term evolution of
stars formed in the \starforge\ simulation suite -- a set of
radiation-magnetohydrodynamic simulations of star-forming turbulent clouds that
include all key stellar feedback processes inherent to star formation. We use
Nbody6++GPU to follow the evolution of the young stellar systems after gas removal.
We use HDBSCAN to define stellar groups and analyze the stellar kinematics to
identify the true bound star clusters. The conditions modeled by the simulations,
i.e., global cloud surface densities below 0.15\:g\,\pcm\,, star formation
efficiencies below 15\%, and gas expulsion timescales shorter than a free fall
time, primarily produce expanding stellar associations and small clusters. The
largest star clusters, which have $\sim$1000 bound members, form in the densest and
lowest velocity dispersion clouds, representing $\sim$32 and 39\% of the stars in
the simulations, respectively. The cloud's early dynamical state plays a
significant role in setting the classical star formation efficiency versus bound
fraction relation. All stellar groups follow a narrow mass-velocity dispersion
power law relation at 10 Myr with a power law index of 0.21. This
correlation result in a distinct mass-size relationship for bound clusters. We also
provide valuable constraints on the gas dispersal timescale during the star
formation process and analyze the implications for the formation of bound systems.
\end{abstract}

\section{Introduction}
Star formation is the result of the complex interplay between physical processes
over a wide range of spatial scales within giant molecular clouds (GMCs) undergoing
gravitational collapse. One key ingredient in this process is turbulence, which is
characterized by large-scale random motions that cascade down to smaller scales
\citep[e.g.,][]{Girichidis2020}. As cloud collapse proceeds, turbulent motions seed
hierarchical fragmentation, leading to stellar groups of thousands rather than a
small number of isolated stars \citep{Pokhrel2018}. Depending on the parent cloud
properties, the star-formation process produces stellar distributions spanning a
broad range of sizes and scales, from massive bound clusters and unbound
associations down to multiple and single-star systems
\citep{Krause2020,Pokhrel2020}.

As star formation models have become more realistic and kinematic observations have
become more detailed, there is increasing evidence that a significant fraction of
stars do not form in bound configurations \citep{Ward2020,Wright2022}, which we
henceforth refer to as \emph{clusters} \citep{Krumholz2019,Chevance2022}. While
most stars in the galaxy may have formed in groups of dozens to thousands of stars
\citep{Lada2003}, the vast majority are currently not in clusters but rather
compose the distributed Milky Way field population. The statistical absence of
relatively old, gas-free star clusters compared to the abundance of embedded star
clusters suggests that most stellar groups dissolve when natal gas from their host
cloud is dispersed due to stellar feedback \citep{Krumholz2019,Krause2020}.
However, observationally determining whether a given group of young stars is truly
bound has historically been challenging given the difficulty of obtaining detailed
kinematic information \citep{Gieles2011}.

Early models of stellar systems using smooth spherical stellar distributions, show
that two key factors inhibit the formation of a bound cluster: low star formation
efficiency (SFE), whereby most of the binding potential is lost with the ejected
gas, and rapid gas dispersal, whereby stars do not have time to adapt their orbits
to the new potential
\citep{Tutukov1978,Hills1980,Adams2000,Baumgardt2007,Proszkow2009,Pfalzner2013,Smith2013}.
Subsequent studies have demonstrated that the introduction of stellar substructure,
motivated by the evidence of filamentary structure in star-forming regions,
produces large variations in the dynamical state of the stars, as the groups
transition from substructured to spherical configurations through dynamical
interactions. Such relaxation produces sub-virial and super-virial dynamical states
that may help or hinder a star cluster's ability to retain members
\citep{Goodwin2009,Lee2016,Farias2018b,Li2019}. 

Furthermore, the structure and dynamics of how the gas interacts with the stars is
such a complex process that most prior theoretical work adopts a simplified and
somewhat arbitrary description for the parent gas cloud or neglects key sources of
stellar feedback, such as radiation pressure, photoionization, and stellar winds;
that affects its dynamics and the subsequent gas dispersal (e.g\
\citealt{Pelupessy2012,Sills2018,Farias2018b}, also see review by
\citealt{Krause2020} ). Consequently, the interplay between these different stellar
feedback processes and the substructures formed by the stars and gas play a crucial
role in setting the stability of young stellar systems and determining whether a
star cluster or association forms. 

The recently developed STAR FORmation in Gaseous Environments (\starforge)
framework \citep[hereafter \citetalias{Grudic2021}]{Grudic2021} has made
significant progress in modeling the evolution of star-forming clouds, including
all relevant forms of stellar feedback. Additionally, since \starforge is capable
of forming individual stars down to $M_\star \sim 0.1~M_{\rm odot}$ it is able to
follow stellar dynamics and the resulting stellar distribution until the natal gas
is dispersed \citep[hereafter \citetalias{Guszejnov2021} and
\citetalias{Grudic2022} , respectively]{Guszejnov2021,Grudic2022}. In the following
paper series,  the \starforge\ group modeled molecular clouds with a range of
initial conditions and studied the role of feedback on the origin of stellar
clustering \citep[hereafter \citetalias{Guszejnov2022}]{Guszejnov2022}, the stellar
IMF \citep[hereafter \citetalias{Guszejnov2022a}]{Guszejnov2022a}, and the
evolution of stellar multiplicity \citep[hereafter
\citetalias{Guszejnov2023}]{Guszejnov2023}. The models explored the impact of
variations in the dynamical equilibrium of the cloud, magnetic field strength,
interstellar radiation field, metallicity, gas density and cloud mass. Altogether
the simulated clouds produce a significant range of stellar distributions, where
most appear to be unbound at the end of the calculation
\citepalias{Grudic2022,Guszejnov2022}. In particular \citetalias{Guszejnov2022}
used the clustering algorithm DBSCAN to identify stellar groups and follow their
dynamical evolution. They found that smaller groups frequently merge to form larger
complexes, which later fragment and expand during 
the post-collapse phase where most of gas is dispersed by stellar feedback.
However, the subsequent evolution and final state of the
stellar groups was not modeled.

In this work, we examine the long-term evolution of the stellar distributions from
the \starforge\ simulations (i.e., once the remaining cloud gas is dispersed) to
characterize their properties and linking these, when possible, to their parent
clouds. 
We follow the terminology convention that a star \emph{cluster} is a group of stars
that are gravitationally bound, while a stellar \emph{association} is a group of
stars that
are not bound but are identified as a single system  \citep[e.g., using a
clustering algorithm, see below,][]{Krumholz2019}. We reserve \emph{group} as a
general term, which encompasses both clusters and associations.
 
In \S~\ref{sec:methods} we describe the numerical codes used in this work and
summarize our methods. We present our analysis and results in \S~\ref{sec:results},
followed by discussion in \S~\ref{sec:discussion}, and conclusions in
\S~\ref{sec:conclusion}.

\section{Methods}\label{sec:methods}
\subsection{\starforge }\label{sec:starforge}
In this paper, we follow the evolution of the stellar complexes formed in the
\starforge\ simulations \citepalias{Grudic2022,Guszejnov2022}. \starforge\ is a
numerical framework developed to model the formation of stars from their parent
Giant Molecular Clouds (GMCs), including the most complete set of physical
processes to date over a wide dynamical range \citepalias{Grudic2021}. These
physical processes include all key sources of stellar feedback such as protostellar
outflows, stellar winds, radiation pressure, photoionization and supernovae.

The simulations are run with the GIZMO code and adopt the Lagrangian meshless
finite-mass (MFM) method for magneto-hydrodynamics (MHD) under the ideal MHD
approximation \citep{Hopkins2015,Hopkins2016a}. Self-gravity is modeled using an
improved version of the Barnes \& Hut tree algorithm \citep{Springel2005a}. The sink
particle orbital integration uses an order four Hermite integrator, allowing the
correct integration of binaries and higher order multiples. Accreting sink
particles have a fixed radius of 18 AU. This radius is also used as a softening
length for close encounters. As sink particles (protostars) form, they follow the
protostellar evolution model from \cite{Offner2009a}. 
Cooling and heating are treated utilizing the thermo-chemistry module from
\cite{fire3}, which includes metallicity-dependent cooling and heating from
$T = 2.7- 10^{10}$\,K, recombination, thermal bremsstrahlung, metal lines, molecular
lines, fine structure and dust collisional processes \citep[see references in][for
details]{fire3}.
Radiative processes are also included, accounting for photon transport, absorption
and emission in 5 bins covering the electromagnetic spectrum. Sources of radiation
include stars (including both the accretion and internal stellar luminosities),
thermal dust emission, and gas continuum and line processes modeled by the cooling
treatment. Other important forms of stellar feedback implemented in \starforge are
stellar winds from massive stars, following the prescription described in
\citetalias{Grudic2021}, and protostellar jets, which are implemented by ejecting a
fraction of the accreted material along the rotational axis of the protostar
\citep{Cunningham2011}. In addition, stars more massive than 8\,\msun\ will undergo
a supernova explosion at the end of their lifetime (at least 3\,Myr), which is
implemented as an isotropic ejection of all mass with a total energy of $E_{\rm SN}
= 10^{51}$\,erg.

\subsubsection{\starforge simulations as initial conditions}

We post-process a selected set of simulations where the fiducial set of initial
conditions is presented in \citetalias{Grudic2022}, and simulations with variations
on the fiducial parameters are presented in \citetalias{Guszejnov2022a}. The
simulations begin with a uniform sphere of gas of mass $\mcl= 20,000\,\msun$, radii
of $\rcl=3-30$\,pc and $T=10$~K; the cloud is surrounded by warm ($T=10^4$~K),
diffuse gas that is 1000 times less dense so that the cloud is initially in thermal
pressure equilibrium with the ambient medium. The cloud turbulence is initialized
with a Gaussian random velocity field with a power spectrum $E_k \propto k^{-2}$
scaled to match the \emph{turbulent virial parameter} (\alphaturb), which
represents the relative importance of the cloud's kinetic energy to gravity,
defined as \citep{Bertoldi1992}:
\begin{eqnarray}
        \alphaturb &= & \frac{5 ||\vturb||^2 \rcl }{3G\mcl},
\end{eqnarray}
where \vturb\ and \rcl\ are the turbulent velocity field and cloud radius,
respectively. As a fiducial value, \citetalias{Grudic2022} set $\alphaturb=2$,
which is characteristic of GMCs in the Milky Way \citep[see
][]{Larson1981,Chevance2022}.

In these models, the importance of the magnetic field relative to the gravitational
energy is parameterized by $\mu$ as:
\begin{eqnarray}
        \mu &= & c_1 \sqrt{\frac{-E_{\rm grav}}{E_{\rm mag}}},
\end{eqnarray}
where $E_{\rm grav}$ and $E_{\rm mag}$ are the gravitational and magnetic energy,
respectively, and $c_1\approx0,42$ normalization constant such that $\mu= 1$
represents a critically stable homogeneous sphere in a uniform magnetic field
\citep{Mouschovias1976}. The fiducial simulation value is $\mu=4.2$ 
i.e.\ $E_{\rm mag} = 0.1E_{\rm grav}$. The simulations include an external heating 
source representing the interstellar radiation field (ISRF), using the default 
assumption of solar neighbourhood conditions \citep{Draine2010}. Dust abundances 
in the clouds assume solar metallicity with a dust-to-gas ratio of 0.01.

Table~\ref{tab:models} summarizes the \starforge\ simulations used in this work.
The \fiducial\ model represents the standard set of parameters used in
\citetalias{Grudic2022} and \citetalias{Guszejnov2022a}, i.e., a
$\mcl=20,000\,\msun$ turbulent molecular cloud that is marginally bound
($\alphaturb=2$) with size $\rcl=10$\,pc and $\mu=4.2$ with local (solar
neighborhood) ISRF conditions and solar metallicity. We also investigate ten
variations of the standard model: low and high turbulent velocity field runs
\alphaL ($\alphaturb=1$) and \alphaH\ ($\alphaturb=4$), respectively; two cases
with 10 (\Bten) and 100 (\Bhundred) times stronger magnetic fields and two models
with ten times higher and lower densities, i.e., clouds with $\rcl =3$\,pc
(\rsmall) and $\rcl=30$\,pc (\rlarge); ISRF increased by a factor 10 (\ISRFten) and
100 (\ISRFhundred); and two models with gas metallicity decreased by a factor ten
(\Zten) and hundred (\Zhundred) relative to the fiducial solar value. 

The global evolution follows the cloud collapse; most stars in the simulation form
at about two initial freefall times. The clouds have an average star formation
efficiency (SFE) on the order of $\sim10\%$. Radiative and wind feedback from the
newly formed massive stars start to disperse the cloud and reduce star formation
around 2 freefall times. Given their high numerical cost, the simulations are
chosen to end after the first supernova goes off. At this point
star formation has mostly ceased and most of the gas was already dispersed from the 
region. Our calculations pick up where the
\starforge\ simulations end (\tedge, see Table~\ref{tab:models}), and we continue
to follow the evolution of the formed stellar groups using the direct $N$-body code
\texttt{Nbody6++GPU} \citep{Aarseth2003,Wang2015} without the gas particles. 
We evolve these stellar distributions
for 200 Myr at which point the stellar groups can be clearly classified as either
clusters or associations.

\begin{table*}
    \centering
    \begin{tabular}{rc cccc c cccc c}
        \toprule
        & & \multicolumn{4}{c}{Cloud initial conditions} & & \multicolumn{4}{c}{Properties at \tedge} & \\
        \cmidrule{3-6} \cmidrule{8-11}
        Model & \# of runs & \mcl & \rcl & \alphaturb & $\mu$ & & $\tedge$ & SFE & LSF & $N_*$ & \texp \\
        &            & [\msun] & [pc] &           &       & & [Myr]    &      &     &       & [Myr] \\
        \midrule
        \fiducial & 3 & $2\times10^4$ & 10 & 2 & 4.2 & & $10.1\pm 0.9$ & $0.080\pm 0.003$ & $0.97\pm0.04$ & $2034\pm131$ & $5.0\pm0.9$ \\
        \alphaL & 1 & $2\times10^4$ & 10 & 1 & 4.2 & & 9.0 & 0.11 & 0.99 & 2464 & $5.0$ \\
        \alphaH & 1 & $2\times10^4$ & 10 & 4 & 4.2 & & 17.1 & 0.039 & 0.98 & 1220 & $7.9$ \\
        \rsmall & 1 & $2\times10^4$ & 3 & 2 & 5.2 & & 3.1 & 0.19 & 0.93 & 4450 & $1.0$ \\
        \rlarge & 1 & $2\times10^4$ & 30 & 2 & 4.2 & & 51.6 & 0.0099 & 0.99 & 358 & $30$ \\
        \Bten & 1 & $2\times10^4$ & 10 & 2 & 1.3 & & 12.3 & 0.078 & 0.91 & 2246 & $7.1$ \\
        \Bhundred & 1 & $2\times10^4$ & 10 & 2 & 0.42 & & 18.2 & 0.056 & 0.92 & 2415 & $7.2$ \\
        \ISRFten & 1 & $2\times10^4$ & 10 & 2 & 4.2 & & 11.2 & 0.092 & 0.99 & 1835 & $6.0$ \\
        \ISRFhundred & 1 & $2\times10^4$ & 10 & 2 & 4.2 & & 9.7 & 0.11 & 0.95 & 2116 & $4.5$ \\
        \Zten & 1 & $2\times10^4$ & 10 & 2 & 4.2 & & 10.9 & 0.045 & 0.15 & 1006 & $5.5$ \\
        \Zhundred & 1 & $2\times10^4$ & 10 & 2 & 4.2 & & 14.8 & 0.022 & 0.76 & 411 & $6.0$ \\
        \bottomrule
    \end{tabular}
\caption{%
        \starforge simulation parameters for the different cloud models, which we
        label as shown in column 1. Second column shows the number of \starforge\
        simulations available for the set, including different initial turbulent
        seeds. Third to sixth columns show the parent cloud initial parameters,
        i.e.\ total mass (\mcl), initial radii (\rcl), turbulent virial parameter
        (\alphaturb) and magnetic field strength parameter ($\mu$), respectively.
        Seventh column shows the end time of the \starforge\ simulations and the
        beginning of the $N$-body modeling, \tedge. The next three columns display
        the cloud properties at \tedge, i.e.,  global star formation efficiency
        (SFE) in column eight, stellar mass fraction within the half mass radius of
        the stars (referred as the local stellar fraction, LSF) in column nine, and
        the total number of stars at \tedge, $N_*$, in the tenth column. Last
        column shows the time in the \starforge\ run where the collapse phase of
        the cloud ends and expansion begins, \texp. Values for the \fiducial\ model
        are averages over the three available simulations in the set.
         }
        
        \label{tab:models}
\end{table*}

\subsection{\nbodycode}

\subsubsection{Numerical Methods}

\nbodycode directly integrates the orbits of stars using a fourth-order Hermite
integrator with a hierarchical block time-step method to increase performance. The
block time-step method works by taking the time-step of individual stars from
pre-defined time-step levels ($l$) with $\Delta t_l = \Delta t_1/2^{l-1}$ based on
how quickly the orbits change.
The calculation cost is reduced by using the \cite{Ahmad1973} neighbour scheme,
which requires gathering a list of neighbours for each particle. Force
updates for neighbours are calculated using smaller timesteps (referred to as
\emph{irregular} timesteps), and for all other stars, forces are updated on a larger
timestep (referred to as \emph{regular} timesteps). 

Accurate integration of hard binary orbits and close encounters requires much
shorter timesteps than those appropriate for the rest of the system and therefore
necessitates special treatment. In contrast to the gravitational softening length
adopted in \starforge, but commonly adopted by most direct N-Body integrators,
\nbodycode\ makes use of the \cite{Kustaanheimo1965} algorithm (referred to as
\emph{KS regularization}). In this scheme a 3-dimensional space describing an
isolated binary orbit is mapped onto a redundant 4-dimensional space on which the
equations of motion are \emph{regular} at collision, i.e.\ when the distance between 
the binary components reduce to zero. The implementation of KS
regularizations in \nbodycode\ uses generalizations of this method for perturbed
binaries and higher order hierarchical multiples \citep[\emph{chain}
regularizations, see][and references therein]{Aarseth2003}. In addition, the code
version we use here employs the GPU parallelization implemented by
\protect\cite{Wang2015}.

\subsubsection{$N$-body initial conditions and models } 
Our procedure following the evolution of the stellar systems begins with the last
\starforge simulation snapshot, at which point stellar feedback has ejected most of
the natal cloud gas. Table~\ref{tab:models} records the time that corresponds to
this last snapshot (\tedge). 
We feed the simulated stellar distributions in the final snapshots into the
$N$-body solver after some additional corrections to account for the different
methods and scopes of the two numerical codes, as discussed next. 

The first challenge is the different treatment of close orbits in both codes. While
\nbodycode does not use any approximation for close encounters and  binary orbits,
\starforge\ uses a softening length that weakens the gravitational potential at
close distances, such that the properties of close binaries found in the
\starforge\ simulations will deform when placed in a regime with no approximations.
Therefore we made corrections to the binary orbits before introducing them in the
$N$-body simulations, as we describe in detail below.

A second complication is that some residual cloud gas remains in the vicinity of
the stars we aim to model. While we restart from snapshots where the gas is mostly
dispersed, the gravitational potential contributed by this gas could still
influence the large-scale evolution of the modeled regions.

In the following sections, we explain how we address these complications during the
initialization of \nbodycode. To summarize our procedure, for each of the
\starforge\ models under analysis, we follow two sets of simulations modeling
distinct scenarios. In both cases, we input the corresponding binary-corrected
stellar distribution into \nbodycode and simulate the evolution for 200 Myr. In the
simulation sets we refer to as \noextpot we evolve the stars after gas expulsion,
assuming complete gas clearance from the cluster, i.e.\ the stars evolve in
isolation. In the second simulation set, which we refer to as \extpot, we evolve
the stars with a gas component remaining within the cluster, which we represent
analytically.

\subsubsection{Binary treatment}\label{sec:binaries}
\begin{figure}
        \includegraphics[width=\columnwidth]{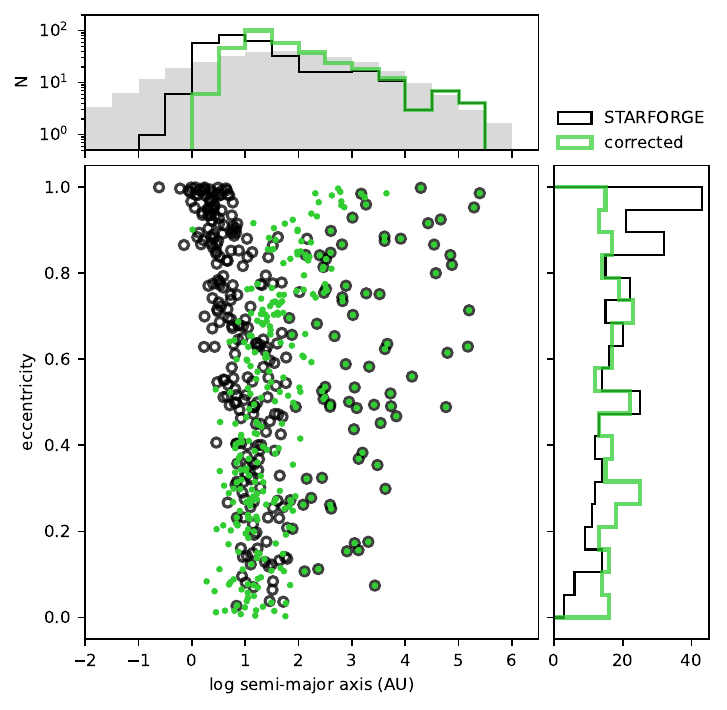}
        \vspace{-.6cm}
        \caption{%
                Semi-major axis versus eccentricity for binaries in the \starforge\
                simulations for the \fiducial\ set at $\tedge=9.1$ Myr (black) and
                after correcting their orbits (green). Under-resolved binary orbits
                produce highly eccentric orbits given their low orbital velocities,
                which is caused by using their velocity orbits under a softening
                length approximation. We correct the orbit of any binary with
                current separation below 60\,\au. The gray shaded area in the top
                panel shows the expected semi-major axis distribution for
                solar-type stars from \protect\cite{Raghavan2010}.
                }
        \label{fig:binfix}
\end{figure}

Binary and multiple systems are a natural outcome of the star formation process
\citep{Offner2022}.  About half of the stars formed in the \starforge calculations
are members of multiple systems, with overall statistics that are in reasonable
agreement with the observed multiplicity and companion star fractions when the
periastron is larger than the softening radius \citepalias{Guszejnov2023}. From a
numerical standpoint, the dynamics of binary stars are challenging to integrate
given their large range of timescales, where the shortest periods are on the order
of days versus the typical dynamical timescale of a star-forming cloud, which is on
the order of millions of years. \nbodycode solves this problem by using KS
regularization routines, however such routines are less advantageous in
tightly-coupled, multi-physics setups, which have additional overheads and lack a
clear separation of timescales between binary and gas evolution. The \starforge
calculations are limited by mass resolution, and equivalently by a minimum scale
length, under which any physical process will be under-resolved, including binary
orbits. \starforge adopts a more practical approach, which uses a conveniently
chosen softening length, to avoid integrating highly expensive orbits on a regime
that is already under-resolved.

The \starforge standard mass resolution  used in all models studied here is
$10^{-3}\msun$. \citetalias{Grudic2021} showed that the smallest Jeans resolved
spatial resolution is of order $\sim20\,\au$, which is adopted as the softening
length. Therefore, any binary with pericenter close to $20\,\au$ is under-resolved,
and we must correct the orbit velocities before evolving these binaries with the
$N$-body solver. We apply the correction to all binaries below a force deviation
tolerance of 10\%, i.e., any pair with current separation below 60\,\au.

Affected binaries have orbital velocities that are too slow, given the weakened
gravitational potential used in the hydrodynamical solver. Consequently, to a
dedicated $N$-body solver, they will appear to have artificially eccentric orbits,
and therefore artificially small semimajor axes. 

We assume that all gravitationally bound stellar pairs produced in the \starforge\
calculations are true binaries, and we correct their orbits by the following
procedure:
\begin{enumerate}
        \item We assign a new eccentricity ($e$), drawn from a uniform distribution.
        \item We retain their current separation but adopt it as the peri-apsis. In
                this way we avoid possible artificial collisions due to highly
                eccentric orbits.
        \item Using the current eccentricity and peri-apsis, we derive the new
                semi-major axis ($a$, which is greater than the original), and
                obtain the magnitude of the velocity at the peri-apsis position.
        \item We find the velocity vector direction by conserving the original
                orbital plane. We obtain the original angular momentum vector, and
                the new relative velocities are assigned such that the direction of
                the angular momentum vector and the center of mass of the binary
                are the same as the original.
\end{enumerate}

Figure \ref{fig:binfix} shows an example distribution of binaries, where  after
correcting the binary orbits, the semi-major axes increase and eccentricities
transform from a skewed distribution peaked at $e=1$ into a uniform one. 

While \starforge also forms triples and higher-order systems
\citepalias[see][]{Guszejnov2023}, most of these are in a hierarchical
configuration. Consequently, only close pairs, which are identified as binaries by
our algorithm, require any orbital correction. We verify that our corrections do
not unbind the wider companions, but otherwise neglect consideration of higher
order systems in our analysis.

Since we are post-processing a binary population derived from resolution-limited
hydrodyamical simulations, close binaries in this work are not an exact
representation of observed populations. However, close binaries represent only a
small fraction of the total observed binary population. For reference, the top
panel of Figure~\ref{fig:binfix} shows the expected distribution of binaries versus
semi-major axis for solar-type stars from \protect\cite{Raghavan2010}. After
correcting the orbits, the discrepancy increases slightly for semi-major axes below
$\sim10~\au$. The observed fraction of binaries with semi-major axes below 10\,\au,
termed the close binary fraction (CBF), is $20\pm2\%$ for Solar-type stars
\citep[see][]{Raghavan2010, Offner2022}. In the same mass range, we obtain a CBF of
approximately $15\%$ in our models. However, note that the corrected orbits do not
include any binaries with semi-major axes below $1~\au$.

A direct consequence of under-representing hard binaries is that the resulting
calculations will underestimate the number of runaway stars at later stages, since
the most energetic ejections depend on the properties of the hard binary
distribution \citep{Perets2012a}. However, as we will see in the following
sections, there is limited dynamical interaction once gas expulsion occurs and the
star clusters expand. Consequently, we expect the primary conclusions of this study
are insensitive to the micro-physics of the binary population.

\subsubsection{Residual cloud gas}\label{sec:backgroundgas}
\begin{figure*}
        \includegraphics[width=\textwidth]{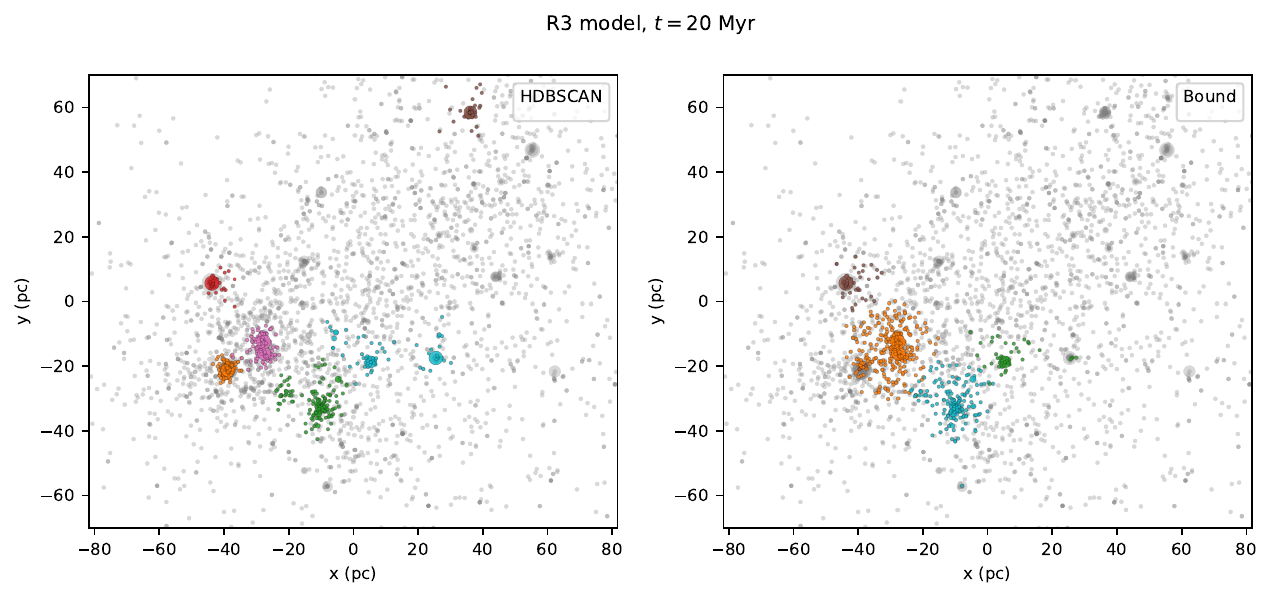}
        \vspace{-0.8cm}
        \caption{Groups identified at 20 Myr in model \rsmall. The left panel shows
                groups identified using HDBSCAN, while the right panel shows bound
                systems (see \S\ref{sec:stellar_analysis}). Each group is assigned
                a different color.
        }
        \label{fig:dist}
\end{figure*}
\begin{table} 
\caption{Gas and star densities within the stellar half mass radius at the end of
        the \starforge\ simulations (\tedge).
}
\centering
\begin{tabular}{rlll}\toprule
Model &$\rho_{\rm gas,0}$        & $\rho_{*,0}$  & $R_{\rm *,h}$   \\
             & \rhounit  &  \rhounit     &    pc           \\
             \midrule
\fiducial    & $1.45\pm 0.9 \times 10^{-3}$  & $0.2\pm0.2$          & $12\pm4$        \\
\alphaL      & $2.43  \times 10^{-4}$        & 0.98                 &5.94             \\
\alphaH      & $2.59 \times 10^{-4}$         & 0.0222               &16.1             \\
\rsmall      & 0.167                         & 2.45                 &5.68             \\
\rlarge      & $6.90\times 10^{-6}$          & $1.73\times 10^{-3}$ &23.9             \\
\Bten        & 0.0135                        & 0.139                &11.0             \\
\Bhundred    & $5.72\times 10^{-3}$          & 0.0664               &12.7             \\
\ISRFten     & 0.0227                        & 0.427                &8.03             \\
\ISRFhundred & 0.0137                        & 0.268                &10.1             \\
\Zten        & $8.41\times 10^{-3}$          & 0.0172               &11.0             \\
\Zhundred    & $3.23\times 10^{-4}$          & 0.0127               &12.7             \\
\bottomrule
\end{tabular}                                                 
        \label{tab:gas}
\end{table}
The \starforge\ simulations we adopt follow the star formation process of star
clusters until star formation has mostly ceased. However, there is still a small
amount of gas that is co-spatial with the stars. To quantify the significance of
the remaining gas, we define the local stellar fraction (LSF) as the ratio of the
total stellar mass to the total mass within the half mass radius of the stars: 
LSF$~= M_{\rm *,h} / (M_{\rm *,h} + M_{\rm gas, h})$.
We find that in most models the LSF is above 90\% (see Table~\ref{tab:gas}). While
we expect its impact to be small, in principle background residual gas could
provide enough binding energy to affect the boundedness of the stellar groups. To
determine if it is indeed a negligible factor, we perform a second series of
simulations with a background gas representation. Since the hydrodynamical gas
structure is complex and difficult to realistically describe in pure $N$-body
simulations, we represent the gas gravitational potential using an approximate
analytical distribution. 

An analysis of the gas distribution indicates that, in most cases, a uniform
distribution is a  reasonable representation for the radial profile. We model a
uniform distribution by adapting the Plummer sphere model implemented in
\nbodycode. This is modeled by adding the gravitational potential generated by a
Plummer density distribution to the stellar potential: 
\begin{eqnarray}
        \rho_{\rm pl}(r) &= \rho_c
        \left(1+\dfrac{r^2}{R_{\rm pl}^2} \right) ^{-\frac{5}{2}},
\end{eqnarray}
where $\rho_c$ is the central density and $R_{\rm pl}$ the scale radius. We choose
a large $R_{\rm pl}$ such that the gas is uniform within the stellar half mass
radius $R_{\rm *,h}$. The central density is the average density within the half
mass radius of the stars. We show the values of the central densities in Table
\ref{tab:gas}.

Since we expect stellar feedback from the stars to continue to disperse the
remaining gas, we estimate the gas gravitational potential decay timescale by
measuring the gas velocity dispersion inside the half mass radius of the stars.
This depletion timescale is given by $\tau_{\rm M} = r_{\rm h}/\sigma_{\rm gas}$.
We assume the gas mass of the Plummer sphere decays exponentially over this
timescale as \citep{Kroupa2001}: 
\begin{eqnarray}
        \label{eq:gasmass}
        M_{\rm gas} &= M_{\rm gas}(0) e^{ - (t-t_{\rm D})/\tau_{\rm M} },
\end{eqnarray}
where $t_{\rm D}$ is a delay time for gas depletion that we set equal to zero,
i.e.\ gas depletion begins at the start of the N-body calculation.

\subsection{Stellar Analysis}\label{sec:stellar_analysis}

\subsubsection{Identifying Groups with HDBSCAN} 
Star formation simulations often form not only one but several stellar groups that
move in different directions after formation
\cite[e.g., ][]{KirkOffner2014,LiKlein2018}. \citetalias{Guszejnov2022} found that
many of the smaller groups merged during formation to become larger groups. In this
work, we focus on the later evolution of the groups analyzed in
\citetalias{Guszejnov2022}, and we study their kinematics, evolution and boundedness
over the subsequent 200 Myr.

We identify groups in the simulation snapshots using the Hierarchical Density-Based
Spatial Clustering  of Applications with Noise (HDBSCAN) algorithm
\citep{Campello2013,McInnes2017,Malzer2020}, implemented in Python\footnote{%
        \href{https://pypi.org/project/hdbscan/}{
                https://pypi.org/project/hdbscan/}
        }. 
HDBSCAN is an extension of the DBSCAN algorithm \citep{Ester1996} that
adopts an adaptive rather than a fixed characteristic size for groups. Both
methods have been previously used to identify stellar groups in surveys 
\citep[see e.g.,][]{Castro-Ginard2018,Hunt2021,Kerr2021,Tarricq2022}.

The simplier DBSCAN algorithm works by identifying points that are ``close" to each
other according to some predefined distance metric, typically the Euclidean
distance.  DBSCAN requires two user-defined parameters: a characteristic minimum
separation scale $\lambda$ and a minimum group size $N_{\rm min}$. A group is
defined to be all stars located within a distance $\lambda$ of another star, where
the total number of connected stars is at least $N_{\rm min}$. Stars that are not
within $\lambda$ of any other stars or who are part of a group smaller than $N_{\rm
min}$ are not assigned to any group.  

DBSCAN works well for single snapshots and relatively well for time-series
where the groups are of a similar size \citepalias[although some extra measures are
still needed to eliminate noise between time-series, see][]{Guszejnov2022}.
In our application, however, the expanding nature of the regions complicates the
choice of $\lambda$. Groups from the same simulation may expand at
different rates depending on whether they are more or less bound, and consequently,
we need to find groups over a wide range of evolving densities. The HDBSCAN
algorithm, which is a generalization of DBSCAN, addresses this specific problem. 

The purpose of HDBSCAN is to find groups on a wide range of scales. Essentially,
HDBSCAN analyzes all possible solutions of DBSCAN for a given value of $N_{\rm
min}$, i.e., all possible choices of $\lambda$, and returns the groups that are the
most persistent over a range of scales.
HDBSCAN requires only two user parameters: $k$, which specifies the number of
neighbors to consider when forming groups and $N_{\rm min}$, the minimum number of
points in a group.

The HDBSCAN algorithm proceeds as follows: First, it uses the specified distance
metric (Euclidean in our case) to find the distance to each star's $k$-th nearest
neighbour. This distance defines each star's $k$-neighbour radius ${\cal
R}_k$.\footnote{%
        Note that the original works termed this distance as ``core-radius''.
        However, we use different term in order to avoid any confusion with
        astrophysical concepts.
} 
Then, it uses the $k$-neighbour radius to define the \emph{mutual reachability
distance} metric, defined by:
\begin{eqnarray}
d_{\rm reach}(a,b) &= \max\{ {\cal R}_k(a),{\cal R}_k(b), d(a,b)\}.
\end{eqnarray}
This definition provides robustness against outliers, so that 
sparsely distributed data, points with a large ${\cal R}_k$, are separated from the
rest of the data by at least ${\cal R}_k$. This avoids the problem that
a small number of data points may act as bridge between two well-separated groups.

Using $d_{\rm reach}$ as a metric, HDBSCAN creates a minimum spanning tree (MST) of
the distribution, from which it constructs a hierarchical tree of connected points.
It then walks the hierarchy by using different distance thresholds, from the large
to the small scales, recording the groups that appear, split, and lose members as
the $d_{\rm reach}$-threshold become smaller. The algorithm then evaluates which
groups \emph{persist} over different scales and selects the most stable
groups \citep[see][for details on how stability is
evaluated]{Campello2013}. 

We use the same input parameters for all $N$-body simulations to select groups. We
adopt a group size $N_{\rm min}=30$ and $k=5$ nearest neighbours to obtain ${\cal
R}_k$. Note that HDBSCAN is equivalent to the DBSCAN algorithm, if instead of
walking through the tree, we use a single $d_{\rm reach}$ threshold to obtain the
groups.

\subsubsection{Tracking Stellar Groups}
Although we adopt the same fixed parameters for all simulations and snapshots,
applying HDBSCAN still produces noisy results between some snapshots, i.e., the
group membership fluctuates. This is because small variations in the stellar
distribution can produce large variations in the assigned group membership. To
address this issue, we down-sample the $N$-body simulation time outputs to 200
equally spaced times and apply HDBSCAN to the closest snapshot to each time. 

Since groups may lose members, merge, split or disappear between successive
snapshots we also develop an algorithm to match groups. Our algorithm builds and
compares two  lists of groups in consecutive snapshots. We call the first group,
the one at the earlier time, the \emph{parent} group and the one at the subsequent
time is  a \emph{child} group. We match each list as follows: 

\begin{enumerate}
        \item We assign a name ${\cal N}_i$ to each parent.
        \item We then compare each child group with each parent. If the child has
                any members of the parent ${\cal N}_i$, then we check if another
                child was already assigned to the parent. \\
               \indent \indent  $\bullet$ If not, we give the
                child the parent's name.  \\
             \indent \indent  $\bullet$    If the name was already assigned to
                another child, then we give the name to the one containing the most
                members and assign no parent to the smaller group (temporarily).
        \item At the end of the list comparison, we assign a new name to
                all unassigned groups. These become potential new parents
                for the groups in future snapshots.
        \item If a parent does not have any child group we keep its name and
                membership list for comparison in future snapshots. Occasionally, a
                group that disappears will reappear in a later snapshot. 
\end{enumerate}

However, even with this procedure, the group membership still fluctuates somewhat
between snapshots as some groups undergo multiple splits and mergers while other
groups are very short-lived and thus are irrelevant to our analysis. Therefore, in
order to have well-defined time-stable groups we require an additional step to
clean the group history. We use a similar method to \citetalias{Guszejnov2022}, by
producing a ``history'' for each star, which we clean in the following way:

We follow the history of each star and identify instances where a star has been
assigned to two or more groups within a characteristic time $\tclean=5 \,\rm Myr$.

\begin{figure*}
\includegraphics[width=\textwidth]{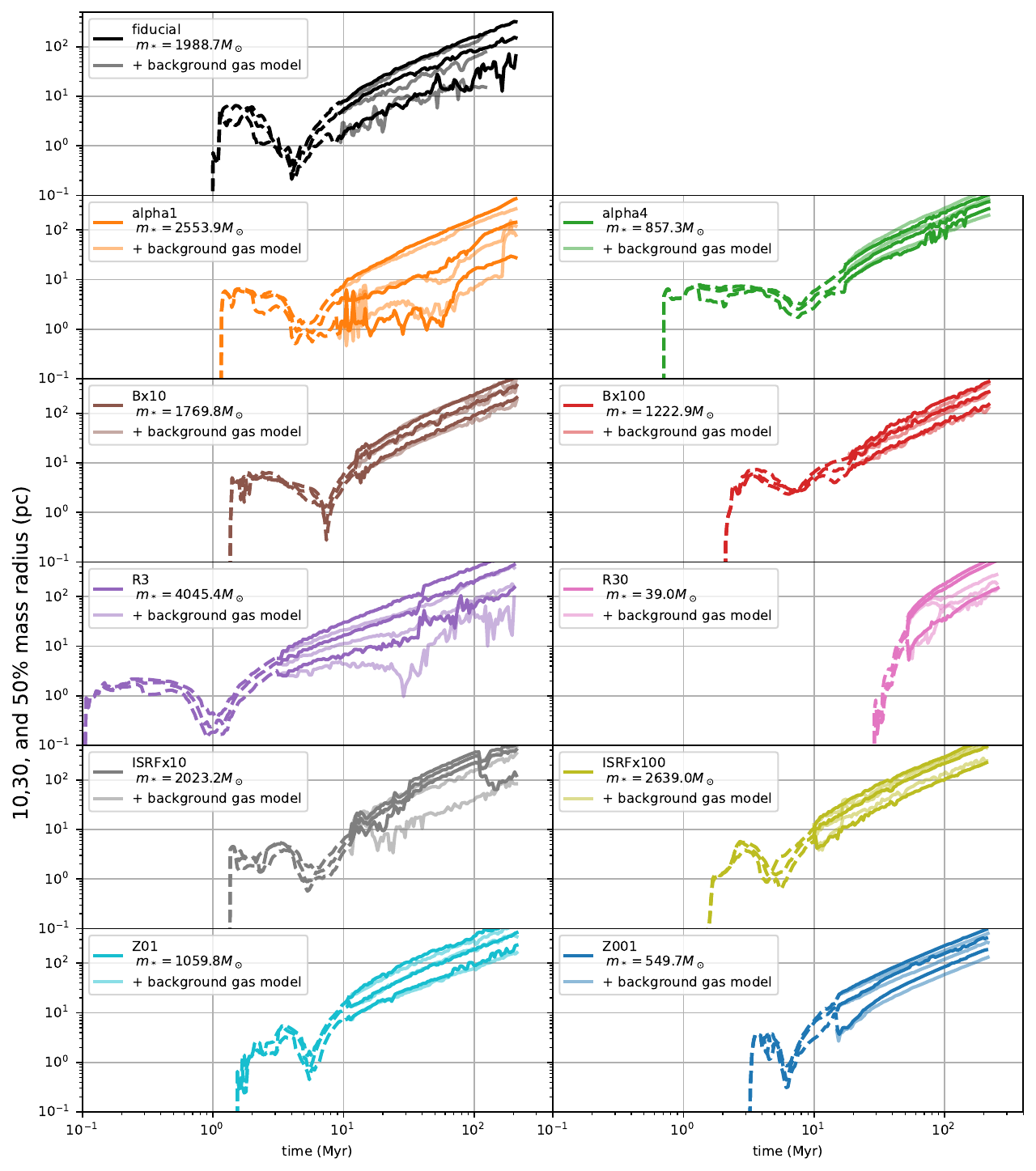}
\caption{Evolution of region size expressed as the 10, 30 and 50\% mass radius
        for each model. Dashed lines show the evolution during the \starforge
        simulation, while solid lines show the evolution in the $N$-body
        simulation. Bold solid lines represent models in the gas-free case
        (\noextpot models), i.e., without any model for background gas. The
        semitransparent lines indicate the evolution including a background
        potential that represents residual gas (\extpot models). 
        }
\label{fig:revol}
\end{figure*}

\begin{figure*}
\includegraphics[width=\textwidth]{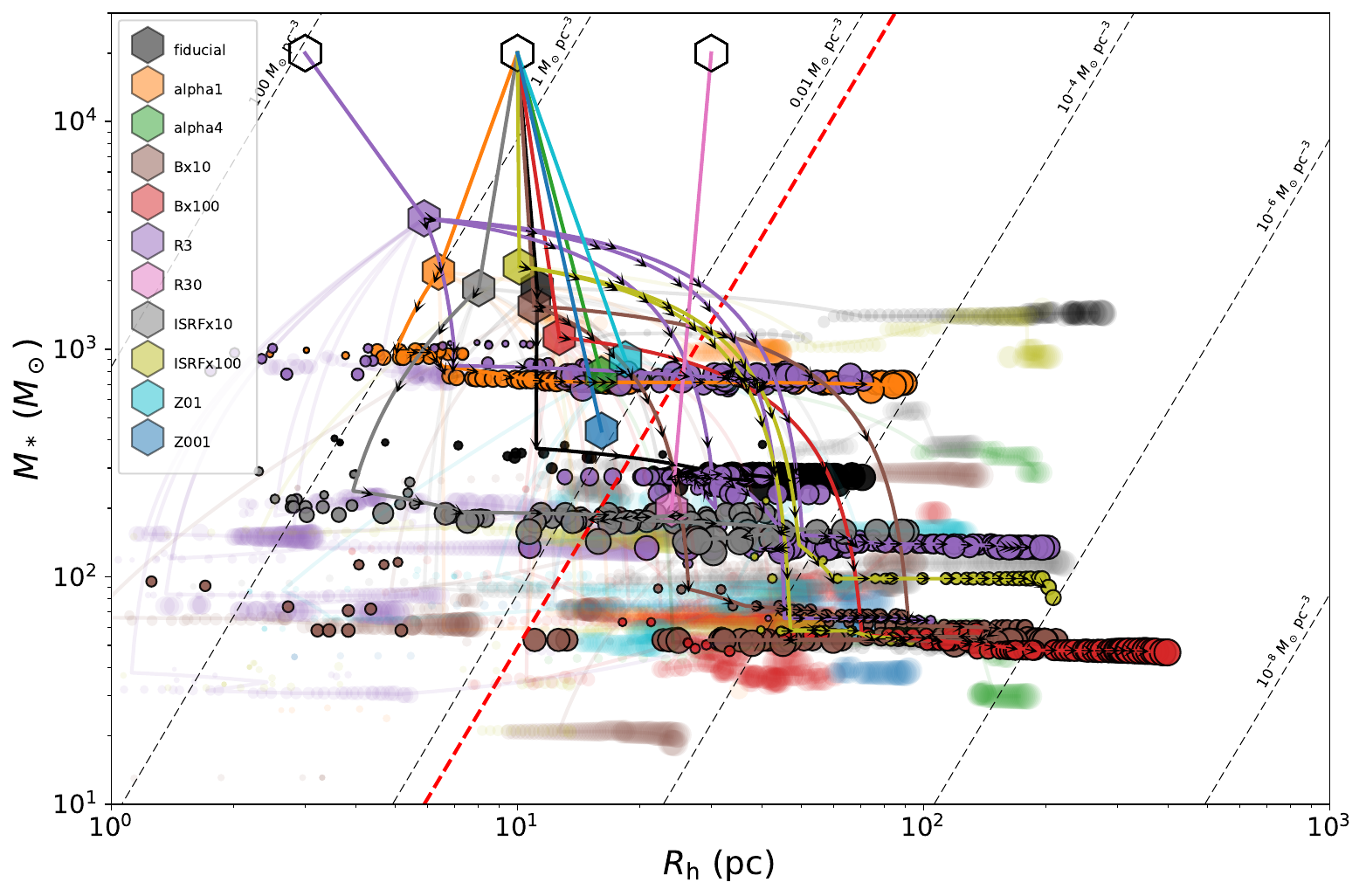}
\caption{%
        Stellar mass versus half-mass radius for systems formed in the \starforge\
        framework. Unfilled black hexagons show the parent cloud initial
        conditions, connected by a line to a colored filled hexagon that shows the
        state of the respective \starforge\ simulations when gas is expelled and
        the simulation ends. From there the $N$-body simulation begins, and we track
        the groups formed in each simulation. We show the state of the groups at
        each time, where the size of the circles is proportional to the cluster
        age. Semi-transparent points are groups identified by HDBSCAN, and black
        contour circles indicate the subset that are bound. Lines (with arrows for
        bound systems) show the average evolution of each identified bound group. 
        Dashed lines indicate lines of constant mass density. Red dashed line represents
        the Jacobi radii at the position of the Sun.  
        }
\label{fig:groupsdiag}
\end{figure*}

The cleaning procedure is as follows:
\begin{enumerate}
\item We create a history of group membership for each star and examine the history
        for membership changes beginning at the last snapshot.
\item We give each star a single label for the last time span of size \tclean,
        where the label represents the most frequent group assignment during this
        time.

\item We search each prior time for a membership change. If the star is not
        assigned to more than one group during the \tclean\ interval we call it a
        \emph{stable label}. If this label represents a group, we call it a
        \emph{stable group label}, i.e., a stable label can also indicate a
        isolated star, which is not assigned to any group. We use the stable group
        label to identify any large time-spans where a group was not identified by
        HDBSCAN. However, as the groups are expanding and there is little
        interaction between them, we find it useful to keep the group label
        identified during this time to provide continuity in the history. This
        allows us to track the group properties.
        Then, we assign the stable group label during any times of ``missing''
        identification.

\item When a group assignment changes, there are three possible cases: \\
 \indent \indent  {\it Case 1)} The
        new group label is the same as the last stable group assigned to the star.
        In this case, we label the star at times in between with the new label,
        ignoring the changes in between.\\
 \indent \indent        {\it Case 2)} The new label is different and the
        old label was not stable, so we assign the last stable label to the time in
        between (either a group or no membership). \\
         \indent \indent  {\it  Case 3)} The label is different
        and the previous label is stable, then we keep the label during this time
        as if no change in membership had occurred.

\item  If the last stable label assigned was a group, then we update the ``stable
        group label`` and record its position.

\item We continue with this procedure until the first snapshot is reached. As a
        final step, we check the groups at every snapshot and remove any groups
        with less than 50 members.
\end{enumerate}

\subsubsection{Identifying stellar clusters}
In addition to applying HDBSCAN, we post-process the resulting groups to identify
their bound members.
In a region with significant substructure, identifying bound systems is not
trivial, as the final result depends on the frame of reference we use. For
instance, two groups may be moving away from each other, such that
selecting one of their center of mass velocities as the \emph{system velocity} will
make most of the other system members unbound. Also, measuring the center of mass
of a system, ideally, would require considering only bound
members, but the bound membership list is the information we are trying to obtain.
To solve this circular problem, we use an iterative method that we 
refer to as ``snowballing'' \citep{Smith2013d,Farias2015}. After providing an
initial position and radius ($r_s$), the method works in two steps: \\
\indent 1) An iterative
process measures the bound members inside the starting radius. Using the center of
mass velocity of stars inside $r_s$, we remove any unbound stars and correct the
center of mass velocity. We repeat this step until it converges to a fixed number
of bound stars. \\
\indent 2) With a robust bound region identified, we now consider all
stars, adding any bound stars to the final sample. After correcting the center of
mass velocity, we repeat this step until the solution converges, which is defined
to be when no more than two stars change membership between steps or a maximum of
100 iterations is reached. 

For the starting radius, $r_s$, and starting position, we use the groups selected
by HDBSCAN, where $r_s$ is the minimum radius containing all stars and the starting
position is the group center. HDBSCAN does not select stars that are too far from
each group and therefore the effective size of the HDBSCAN group is a reasonable
choice for $r_s$.

Figure~\ref{fig:dist} shows the groups identified with HDBSCAN (left) and the bound
groups identified with algorithm described above (right) for model \rsmall. Of the
seven groups found with HDBSCAN, only two contain a significant number of bound
members. Note that even though we use each of the HDBSCAN groups as a starting
point to search for bound stars, some of these individual groups are bound and our
method merges them together. Although we identify groups using HDBSCAN and adopt
these groups as seeds for our cluster identification procedure, throughout this
paper we analyze both sets of groups independently. The HDBSCAN groups represent
how observers identify groups; sufficiently accurate kinematic information is not
available in most cases to determine whether the groups are truely bound.
The bound groups we identify using the second procedure represent true star
clusters. The first set are useful to compare with and interpret observations,
while the latter help us understand the physics behind star cluster formation.

\section{Results}\label{sec:results}
\begin{figure*}
        \includegraphics[height=6.3in]{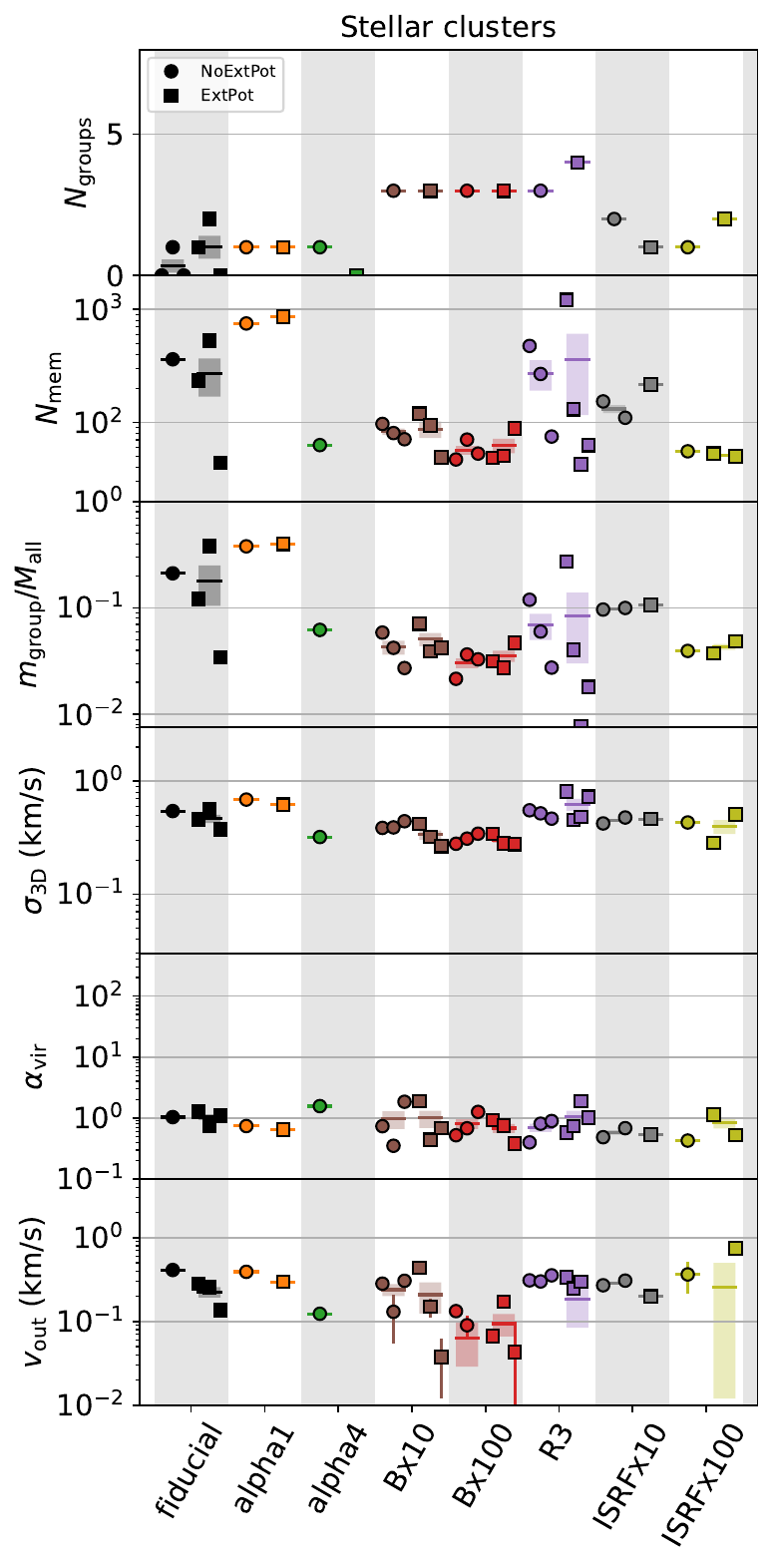}\hfill
        \includegraphics[height=6.3in]{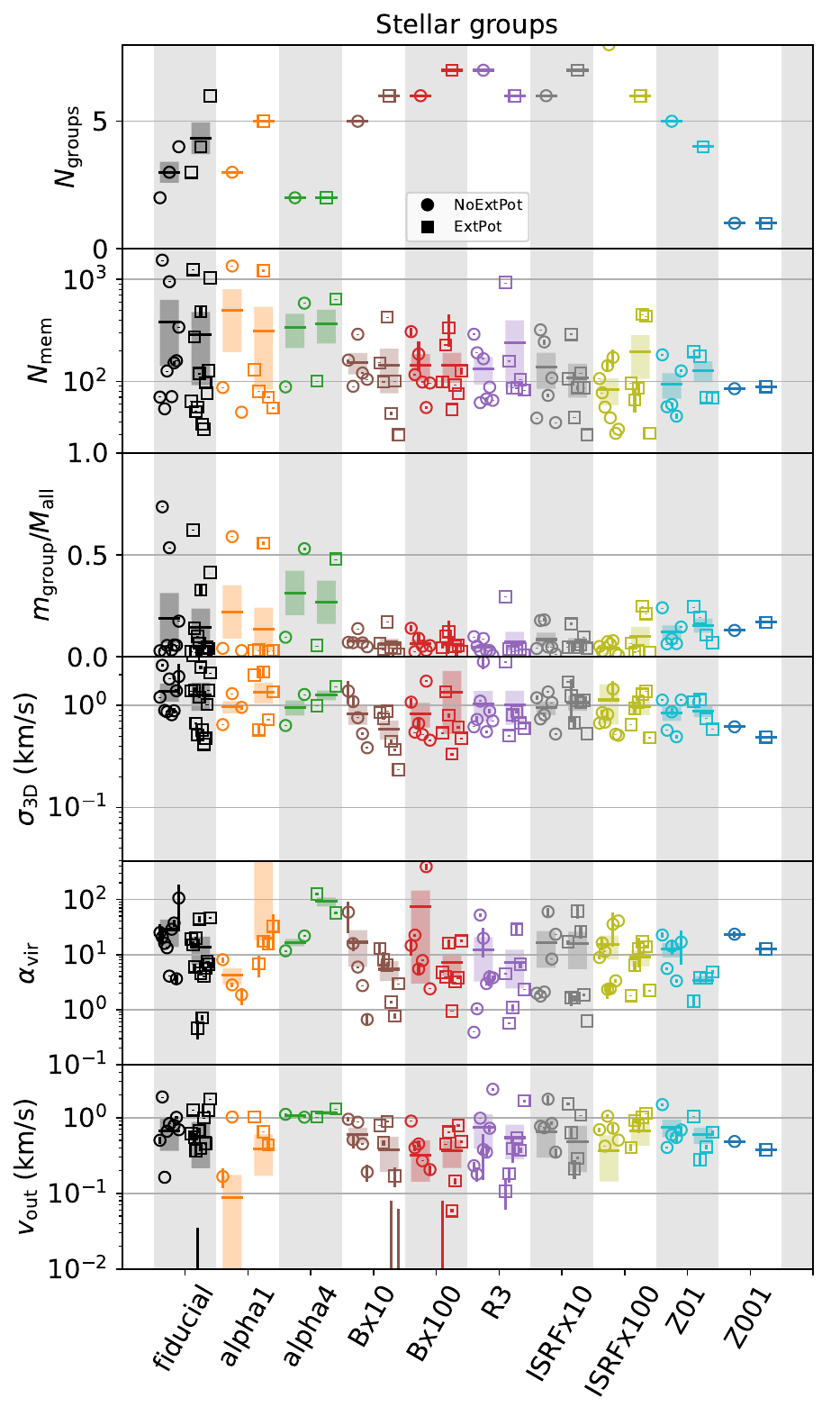}

        \caption{ 
        Time-averaged properties for groups of stars detected in each model between
        8 to 12 Myr after the region's expansion. Left panel shows the results for
        bound systems while left panels shows the results for systems found using
        HDBSCAN. Circles denote groups in models with no background gas
        (\noextpot), while squares denote groups in models that account for
        background gas (\extpot). Note that the \fiducial\ set shows the results of
        three different runs. From top to bottom: the first panel shows the number
        of identified groups in each model, the second panel shows the number of
        members in each group, the third panel shows the mass of the groups
        relative to the total stellar mass, the fourth panel shows the 3D velocity
        dispersion of the stars, the fifth panel shows the virial parameter of each
        system, where $\alpha_{\rm vir} = 1$ represents virial equilibrium, and the
        bottom panel shows the median expansion velocity from the center of each
        group.
        }
        \label{fig:grouppar}
\end{figure*}
\subsection{Evolution and survival of star clusters}

In this section we present the long-term evolution of the stars born in the 
\starforge\ simulations. We compare the evolution with and without a background
potential that represents residual gas. Figure~\ref{fig:revol} shows the global
evolution of the 10, 20 and 50\% radius relative to center of mass of all stars in
each simulation, where the \starforge\ evolution phase is denoted with dashed lines. 
The evolution of the young stars, modeled by \starforge, shows that star
formation happens over a few to 10 million years. As collapse proceeds, the stars
become more centrally condensed (in all but the R30 simulation) at which point
stellar feedback is strong enough to stop star formation and disperse the gas.
In all runs the stars and residual gas do not have enough gravity to confine all the
stars; the global SFE is very low, and, consequently, the stellar complex expands.
Therefore, each GMC produces the equivalent of an association complex. As
previously shown in \citetalias{Guszejnov2022}, this includes a range of smaller
groups. We analyze the behavior of the subgroups in \S\ref{sec:groups}.

We find the half mass radii of all regions expand monotonically, indicating that
the systems are not bound at this scale. However, for the 30\% and 10\% mass
radius, two stellar complexes show signs of dynamical evolution. Model \alphaL\
shows a relatively constant 30\% mass radius for about 50~Myr, before starting to
expand again. This is a sign that a considerably large bound system was formed (see
\S\ref{sec:groups}). However, large formed star clusters do not always show a
signature in the size evolution of the region. As we show below, one of the
\fiducial\ models produces a similar size star cluster, but the global evolution
of the region still shows monotonic expansion. 

In general, accounting for the residual gas does not greatly affect the behavior of
the expanding regions. However, in some particular cases, adding a background
potential triggers the capture of a considerable number of new members. In one
extreme case in the \rsmall\ model, background gas triggers the capture of
$\sim500$ new members, by merging two big clusters (see below). This affects the
overall size of the region as the most prominent clusters merge into one rather
than separating apart.
In another notable case, the \alphaL model forms a single cluster that is slightly
more massive than in the \rsmall\ model. The addition of a background potential
causes only a minor increase in the cluster's mass. However, the evolution of the
half-mass radius exhibits distinct oscillations over time between the two cases.
This reflects ongoing dynamical evolution that is absent in other expanding
regions with less massive clusters. These instances underscore the fact that the
intrinsic chaotic nature of the N-body problem implies that the inclusion of a
background potential, even one representing a small mass fraction, may trigger
unexpected behaviors, particularly in substructured stellar regions.

\subsection{Characterizing stellar groups}\label{sec:groups}
One important motivation of this work is to characterize the populations of star
clusters that form in the \starforge\ calculations. However, there are two
approaches we can take for this problem. One, is to characterize the groups of
stars that observers would detect by using clustering algorithms such as HDBSCAN.
The other approach is to use the full available information to characterize the
true bound systems that form in each model. Both approaches provides useful
insights into the star cluster formation process, and therefore we will present
both.

Figure~\ref{fig:groupsdiag} presents an overview of the identified clusters and
groups and their mass-size evolution. This diagram illustrates that, in general,
groups and clusters initially expand without losing a significant number of
members. We also include the Jacobi radii at the Solar position as a reference, and
we observe that most groups fill these radii with their half-mass radii soon after
formation. This suggests that most of these groups would be stripped out in a
galactic environment shortly after formation. Therefore, our analysis is primarily
constrained to the first 10 million years after their expansion begins.

\subsubsection{Groups sizes and frequencies}
Figure~\ref{fig:grouppar} shows a comparison of different parameters for each group
averaged over a narrow time-span after the expansion of the region, early enough
for the regions to be young and late enough for the groups to be well separated,
i.e., we choose to average the properties between 8 to 12 Myr after the expansion
of the region begins.
The left panel shows the
results for bound systems, while the right panel shows groups identified with
HDBSCAN.  Squares indicate the models that represent residual gas using a
background potential, the \extpot\ case, while circles indicate models that neglect
residual gas, the \noextpot case. 
Note that the cases with and without background gas are statistically similar in
all cases, although there are slightly more massive systems in the \extpot case. 

While HDBSCAN identifies on the order of 2 to 8 groups in each simulation, we see
that a smaller number are identified as (bound) clusters. The largest numbers of
clusters are found in the strong magnetic field models \Bten, \Bhundred\ and the
high-surface density model \rsmall, which each have three to four bound systems.
However,  the clusters in the stronger magnetic field models are relatively small
and contain less than a hundred members each, which is smaller than the typical
cluster size in the \fiducial\ case. These small clusters do not affect the half
mass radius of the stellar complex (see Figure~\ref{fig:revol}), as they contain
less than  10\% of all the stars and are moving away from each other. The stronger
magnetic field induces the formation of a larger number of bound systems, but their
total mass is still less than the total cluster mass in the \fiducial\ case and, as
we will see below, most of them are short lived.  

The largest star clusters are found in the \rsmall\ model where the size of the
clusters varies strongly with the addition of background gas. In the \extpot\ case,
the largest cluster contains $\sim$1100 members. It is accompanied by three smaller
clusters with $\sim$30, 50 and 100 members. In the \noextpot\ case the largest
clusters have $\sim500$ and $\sim200$ members. The big difference in membership
between the \extpot\ and \noextpot\ models is caused by the merging of the two
larger star clusters formed in the \rsmall\ model. The mass of both clusters
together triggers the capture of additional members increasing the difference
between the \extpot\ and \noextpot cases. The bigger cluster contains about 28\%
(see below) of the stars in the simulation and triggers a different evolution in
the 30\% radius of the region as we discuss above.

The next largest cluster is formed in the \alphaL\ case, which has a single cluster
with $\sim800$ and $\sim$900 members in the \noextpot\ and \extpot\ cases,
respectively. The higher turbulent velocity case, \alphaH, produces one ($\sim60$
members) or zero bound systems. 

In the low-density cloud model, \rlarge, no
clusters or groups were identified due to the limited number of stars formed and
their dispersed distribution within the simulation. Compared to the formation
event, specifically the gas expulsion timescale (see
\S~\ref{sec:boundfraction}), the dynamical timescale of the region is
relatively long, which hampers any interactions between stars or adjustments of
their orbits to the relatively rapid decrease in the region's gravitational
potential.

The third row of panels in Figure~\ref{fig:grouppar} shows the fraction of stellar
mass in each group relative to the total stellar mass. For individual star
clusters this represents their bound fractions, while the total mass in star
clusters represents the total bound fraction of the stellar complex, which we
discuss in \S~\ref{sec:boundfraction}. Only the \fiducial, \rsmall, and \alphaL
models form star clusters containing more than 10\% the stars in the region. The
\alphaL\ model has the highest individual bound fraction, $\sim40\%$, by a
slim margin.

The right panel of Figure~\ref{fig:grouppar} shows that HDBSCAN identifies more
groups and that these groups are in general bigger than the identified
clusters, as energy constraints are not taken into account. The \fiducial\  and
\alphaL\ cases have the largest groups with more than 1000 members. HDBSCAN finds
some small unbound groups in the models with lower metallicity,
which formed no clusters. This indicates that these runs still contain
substructure, although all
the groups are smaller than 300 members with less than 100 members total in the
\Zhundred\ case. Lower metallicity also makes it more challenging to
form groups: fewer stars form overall because the increased HII region temeperature
and reduced recombination greatly increases the rate of HII region expantion. 

\begin{figure*}
\includegraphics[width=\columnwidth]{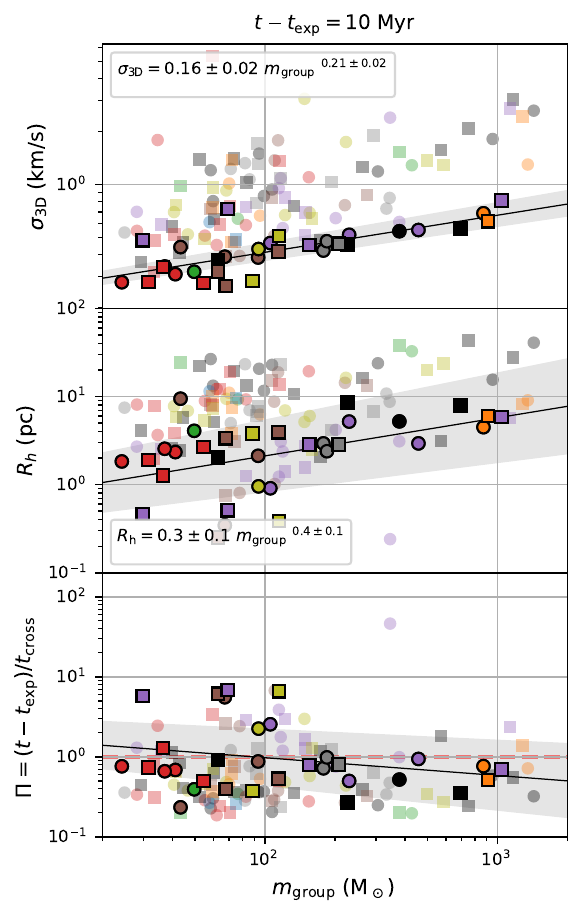}
\includegraphics[width=\columnwidth]{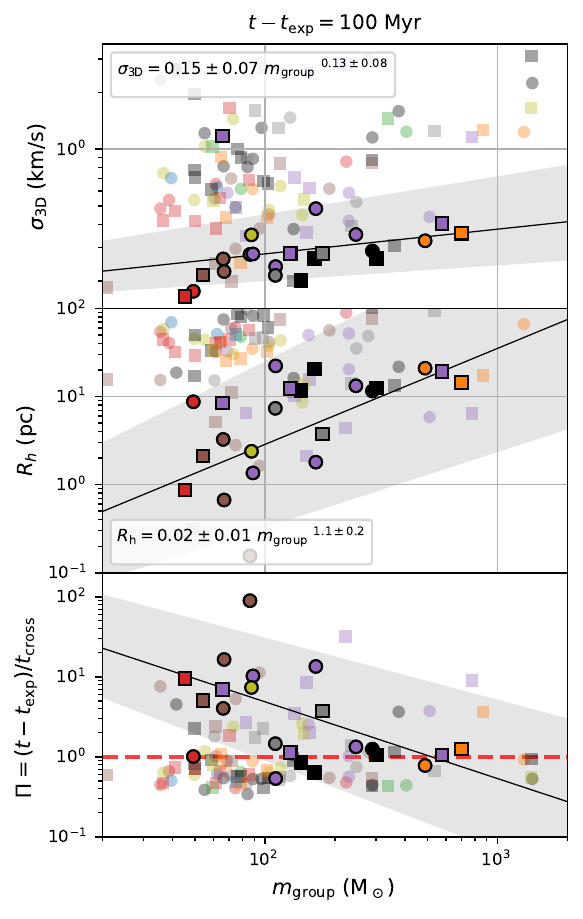}
\caption{\label{fig:alphamass}
        Kinematic parameters for stellar clusters (solid symbols) and groups
        (semi-transparent symbols) as a function of \mgroup. Values are measured at
        10 Myr (left panel) and at 100 Myr (right pannel) after the region
        expansion. Circles and squares show the \noextpot\ and \extpot\ cases,
        respectively. Top panels display the three-dimensional velocity
        dispersion, $\sigma_{\rm 3D}$, middle panels show the group half mass
        radius $R_{\rm h}$,  and bottom panel show their dynamical age, $\Pi$.  
        Bound cluster size and velocity dispersion correlate with group mass
        according to the fit equations shown in the respective panels. The solid lines
        in the bottom panel show the above relations applied to the dynamical
        time. The red dashed lines  $\Pi =1$, which has been proposed as a
        threshold between star clusters and associations \citep{Gieles2011}. 
        }
\end{figure*}

\subsubsection{Dynamical state and scaling relations  }
\label{sec:scaling}

The fourth row of panels in Figure~\ref{fig:grouppar} shows the three dimensional
velocity dispersion of the clusters and groups, $\sigma_{\rm 3D}$.
We remove orbital motion from the velocity dispersion by adopting the center
of mass velocity for binary pairs; consequently, the dispersion provides a direct
measure of the internal dynamics. Interestingly, all star clusters have a similar
velocity dispersion that is independent of the parent cloud properties. The
cluster velocity dispersions range from 0.2 to 0.8\,\kms, which is well
below the bulk simulation gas velocity dispersion. Although the HDBSCAN group
velocity dispersions are higher (0.2-20\,\kms), they also do not appear to
correlate with the parent cloud kinematics. 

We obtain a similar result when measuring the group virial parameter (see fifth row
of panels). By construction all star clusters have $\alphavir\sim1$, while
the HDBSCAN groups span a wider range with $\alphavir \sim 1 - 100$. This suggests
that many observed young groups identified by proximity rather than kinematics may
be significantly unbound.

While we find no correlation between the cloud initial conditions and the dynamical
state of the groups, we do find a scaling relation between velocity dispersion,
cluster size and the stellar group mass, \mgroup, as illustrated in
Figure~\ref{fig:alphamass}. The top panel shows a tight scaling relation between
velocity dispersion and cluster mass. 
Using the least squares method, we fit a power law to the clusters at 10 Myr after
the cluster expansion (\texp, see last column of Table~\ref{tab:models}), resulting
in the relation $\sigma_{\rm 3D} \propto \mgroup^{0.21\pm0.02}$. We find a similar
power law exponent at 100 Myr, $0.13\pm 0.08$ (see the right panel of
Figure~\ref{fig:alphamass}).
Note that in general, we do not expect the power law index to remain static over
time, because the velocities of bound systems change due to the evolution of the
multiple systems, the ejection of members, and stars adapting to the expansion of
the cluster. However, as the clusters analyzed here are relatively small and
undergo relatively little additional relaxation, the relationship between group
mass and velocity dispersion does not vary much. 

The middle panels of Figure~\ref{fig:alphamass} show a broader correlation between
cluster mass and radius, where $R_{\rm h} \propto m_{\rm group}^{0.4\pm0.1}$. This
relation varies significantly with time, such that the power law index rises to a
value of $1.0\pm0.1$ at 100 Myr. These correlations arise as a consequence of the
narrow velocity dispersion mass relation. Since star clusters are bound, their
velocity dispersion, mass, and radius are related through the virial parameter as
follows:
\begin{equation}
        \alpha_{\rm vir} = \frac{2 E_{\rm kin}}{E_{\rm pot} } = 
        \frac{1}{2}\frac{\sigma_{\rm 3D}^2 R_{\rm h}}{\eta G \mgroup},
\end{equation}
where $E_{\rm kin}$ and $E_{\rm{\rm pot} }$ are the kinetic and potential energy of
the stars respectively, $G$ is the gravitational constant, and $\eta$ is a
profile-dependent constant for the potential of the distribution of stars. Using
the half-mass radius as the scale radius, for a Plummer sphere, this value is
$\eta=0.13$. Then, based on the $\sigma_{\rm 3d}$-$\mgroup$ relation above, we
would expect $R_{\rm h} \propto \mgroup^{0.58}$ and $R_{\rm h} \propto
\mgroup^{0.9}$ at 10\,Myr and 100\,Myr, respectively. These values are very close
to the power law fits over-plotted in the middle panels of
Figure~\ref{fig:alphamass}, where variations are due to the structure and different
expansion rates of  the individual clusters.

One way to assess the rate at which star clusters expand, and consequently their
stability over time, is to compare their age to their current crossing time. This
ratio between age and crossing time is termed the dynamical age ($\Pi$). Systems
that undergo rapid expansion exhibit low values of $\Pi$, which decrease as time
progresses. In contrast, stable systems show an increasing dynamical age,
reflecting relatively constant crossing times over time. 
Applying the relationships mentioned earlier, we observe a shift in the correlation
of cluster crossing times over time. Initially, at 10\,Myr, clusters of all masses
follow a rather uniform crossing time, with a correlation of $\mgroup^{0.2}$.
However, this correlation becomes more pronounced and linear by 100\,Myr,
highlighting that more massive clusters expand faster than smaller clusters.

The bottom panels of Figure~\ref{fig:alphamass} show $\Pi$ at 10 and 100 Myr, where
we adopt the time since expansion ($t-t_{\rm exp}$) as the cluster age.
\protect\cite{Gieles2011} proposed a value of $\Pi=1$ as a threshold to distinguish
between star clusters and associations. We observe that larger clusters sustain a
value of $\Pi$ close to unity for over 100\,Myr. In contrast, smaller clusters,
which do not expand as rapidly, increase their dynamical age values over time,
rising from $\lesssim1$ to values closer to 10 at 100\,Myr. However, as these
systems are small and their dynamical timescale is short, at 100 Myr about 40\% of
clusters with masses below 100\,\msun\ dissolve. For clusters with masses above
100\,\msun, this figure drops to 10\% of dissolved clusters.

This result shows that small clusters, while able to form stable systems,
can sustain their increasing densities for only a short period. Larger clusters
form stable configurations, but dynamical interactions, rather than dissolving
them, cause a constant expansion.\\

Stellar groups follow similar trends to those of the clusters but exhibit
significantly more scatter. Like the clusters, their properties do not correlate
with initial cloud properties, and groups formed within super-virial
clouds are not different in radius or velocity from ones formed in sub-virial
clouds. However, high-velocity stars cause most stellar groups to scatter above
the mass-velocity dispersion relation for star clusters. In the mass vs. half-mass
radius diagrams, the groups tend to lie above the relationship observed for
clusters. This is mainly due to the fact that HDBSCAN groups typically consist of a
larger number of members that are more sparsely distributed across larger volumes.
However, this is not always the case.\footnote{ Some exceptions exist as HDBSCAN
        tends to find groups well defined in space. While bound members may be
        located on a wider region and we apply no constraints in the location of
such members. These cases, however, are uncommon. }

The dispersion in velocity and high-mass radius exhibited by the stellar groups
produces significant scatter in the dynamical age vs. mass diagram. Although, the
majority of groups show dynamical ages below 1 at all times, a significant fraction
of them scattered above this threshold.
This variation is more prominent for groups with masses $\lesssim$100\,\msun in
particular. Above this threshold, groups and clusters exhibit similar trends,
possibly attributed to the fact that groups encompass bound clusters. Given their
larger size, a greater proportion of group members are bound. 

In the regime explored in this work, for clusters ranging from 30\,\msun to
1000\,\msun, higher mass clusters expand more rapidly than lower mass clusters. It
remains uncertain to what extent these findings extend to larger systems, as the
increase in cluster mass might mitigate the dynamically induced expansion due to
the deeper gravitational potential.
Nevertheless, these results imply that the stability of newly formed star clusters
is predominantly determined by their mass rather than their initial conditions.

\subsection{Group evolution and kinematics}\label{sec:kinematics}

\begin{figure*}
\includegraphics[width=\textwidth]{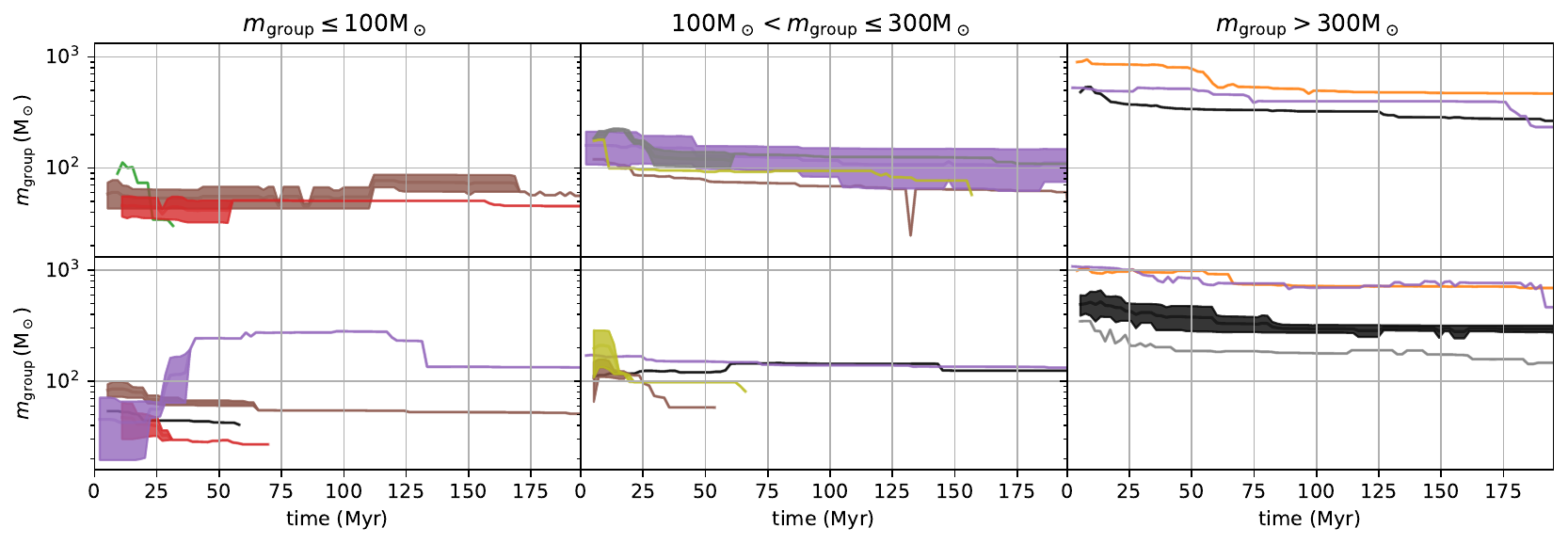} \\
\includegraphics[width=\textwidth]{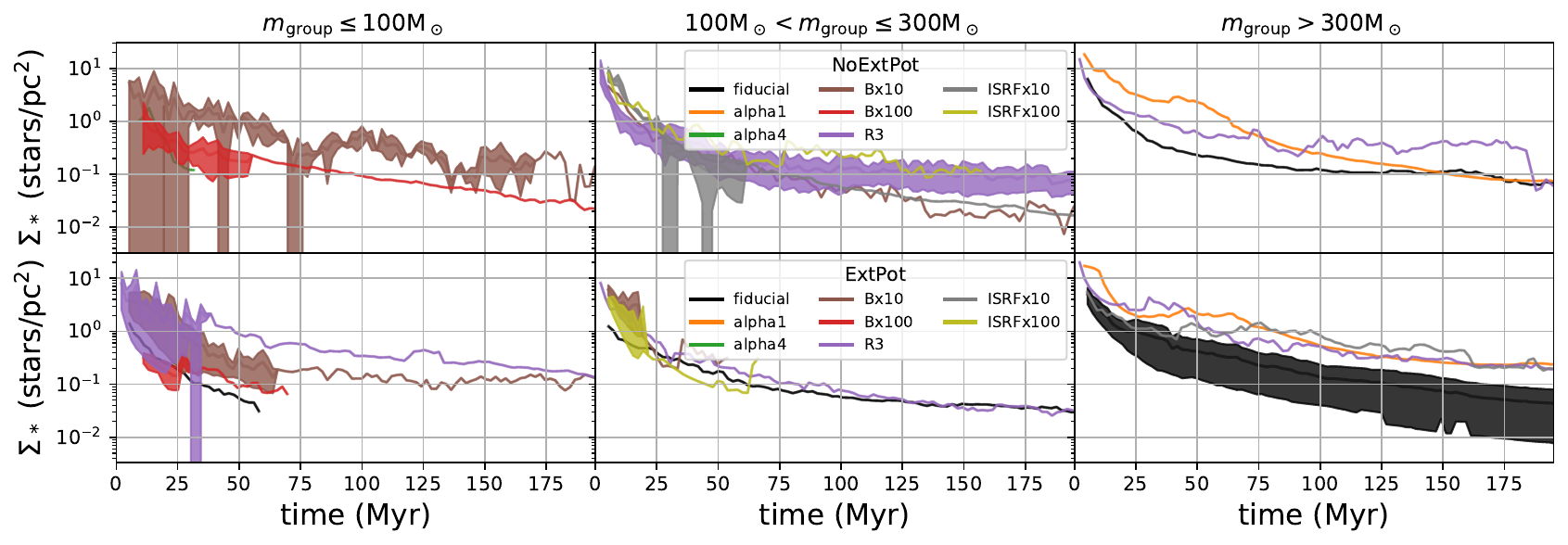} \\
\includegraphics[width=\textwidth]{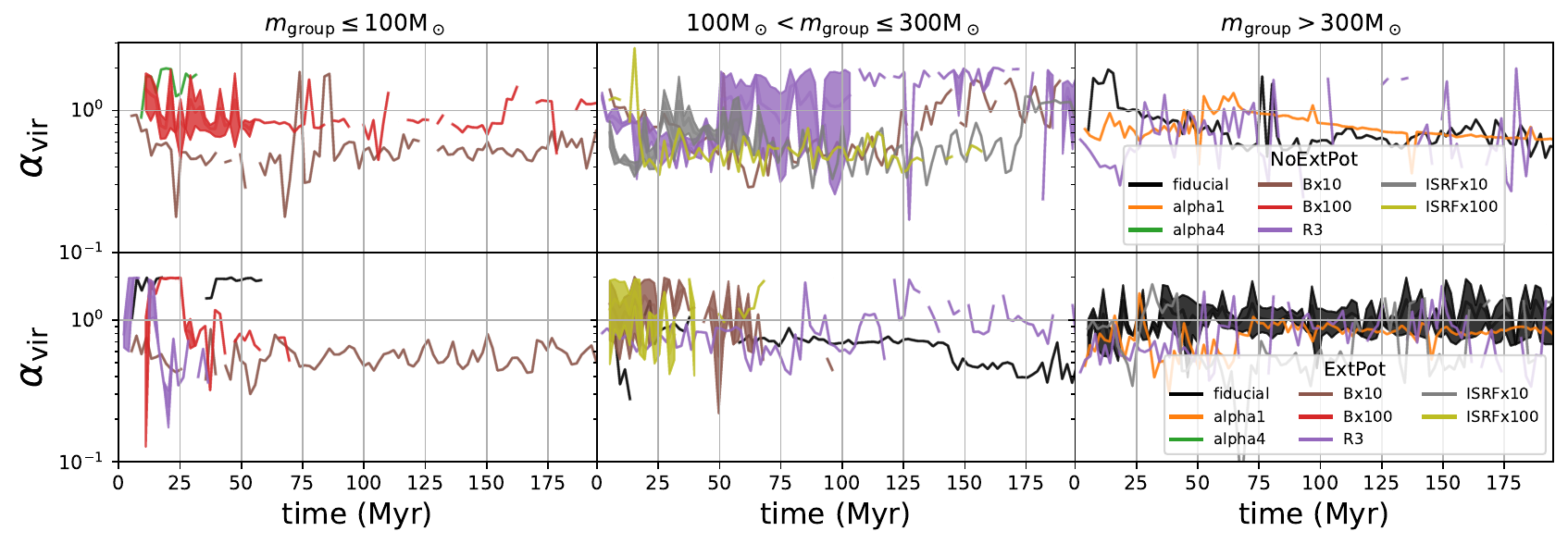} 
\caption{%
        Evolution of different parameters grouped by mass range for identified star
        clusters.   From top to bottom: evolution of the cluster mass (\mgroup);
        cluster surface density ($\Sigma_*$) versus time;  virial parameter
        ($\alphavir$) versus time. In each set of six panels, the bottom and top
        rows show simulations with (\extpot) and without (\noextpot) residual gas,
        respectively. Each line shows the average value for the groups formed from
        the same parent cloud for the given mass range. The shaded areas show the
        standard deviation from the average; lines with no shading represent a
        single group. 
        }
\label{fig:evol_bound} 
\end{figure*}

The clouds analyzed in this work, which have a total initial mass of $2\times
10^4\,\msun$, form groups of stars with masses from $\sim20\,\msun$ to
$\sim1000\,\msun$. To further investigate mass-dependent evolutionary trends, we
group these systems into three mass ranges, where
\emph{small} groups have $ \mgroup  \leq 100\,\msun$,  \emph{medium}  groups have
$100\,\msun < \mgroup \leq 300\,\msun$ and \emph{big} groups have $\mgroup \geq
300\,\msun$. Although the group mass evolves as the membership changes, we adopt
$t=10$ Myr after the expansion begins to characterize the mass of the systems (as
we did for Figure~\ref{fig:grouppar}).

Figure~\ref{fig:evol_bound} shows the evolution of each cluster's mass, two
dimensional stellar surface density ($\Sigma_*$) within each cluster's half-mass
radius, and stellar virial parameter. The lines indicate averages for clusters
formed from the same cloud and in the same mass range, while the shaded areas show
the standard deviation from the mean; single lines represent a single surviving
cluster. About 40\% of small star clusters dissolve before 70 Myr due to their low
mass and shorter crossing times, making them susceptible to dynamical evaporation.
However, around half of them manage to survive for up to 200 Myr. With increasing
cluster mass, their longevity improves, a result attributed to the mass-dependent
cluster expansion discussed in \S\ref{sec:scaling}. Larger clusters primarily
experience mass loss during their early stages when densities are higher. As these
clusters expand, their crossing times increase, reducing the occurrence of
dynamical interactions that could lead to evaporation.

The middle panels of Figure~\ref{fig:evol_bound} show that the stellar half-mass
surface density, $\Sigma_*$, declines during the first $\sim$100 Myr for all
clusters. A few clusters, namely one small cluster in model \Bten, a large cluster
in the \fiducial\ model and a couple intermediate-sized clusters in model \rsmall,
reach a steady-state surface density of $\Sigma_* \sim 10^{-1}$ stars/pc$^2$,
indicating that expansion has halted. As discussed in \S\ref{sec:scaling},
high-mass clusters exhibit more prominent density changes during the initial 100
Myr. By examining the stellar surface densities, we note that low-mass clusters
experience a slower decrease during this time compared to higher-mass clusters that
exhibit a rapid initial decline in density. The overall evolution and final cluster
surface densities appear to be largely insensitive to the initial gas conditions,
however. Similarly, the presence of residual gas seems to have little effect on the
final surface densities.

One prevailing trend is that the virial parameter remains relatively constant over
time. As the star clusters expand, there is minimal dynamical evolution
observed. We do not identify clear relationships between mass and $\alphavir$.
Notably, significant fluctuations in the virial parameter overlap across different
clusters. These variations result from factors such as the precise selection of
members, accurate identification of binary and multiple systems, and the chosen
reference frame for kinetic energy measurement. Slight discrepancies between two
consecutive snapshots could lead to significant alterations in the virial parameter
value, particularly when dealing with a small number of members.

\subsection{Star cluster birth environment}\label{sec:traceback}

In the classical picture of star cluster formation, the fraction of mass that
remains bound after gas removal depends on the global SFE  and also the timescale
on which such gas leave the region. High SFE and long gas expulsion timescales
offer the most favorable conditions to form bound clusters
\citep[e.g.][]{Baumgardt2007,Smith2013}. 
This dependence has been studied mainly by the use of pure $N$-body simulations
with spherically symmetric background potentials to mimic the influence of the gas.
Although in this work we also make use of such potentials, we do so at the later
stages of gas expulsion where the potential generally represents only a few percent
of the mass and its contribution to the evolution of the cluster is minor. The
bound clusters studied here are the result of the complex interactions between
stars and gas, each with their own substructures. 

The STARFORGE simulations provide a unique opportunity to study the early dynamics
between gas and stars in a realistic setup, which can be compared with the
classical picture of star cluster formation. In this section we investigate how the
primordial kinematics of stars within the parent cloud and the cloud evolution
affect the structure and survival of star clusters.

\subsubsection{Gas expulsion timescale}\label{sec:taum}

\begin{table}
        \caption{
                Estimated gas expulsion timescales and fit parameters. The first
                four columns indicate  the model name, estimated gas expulsion
                timescale $\tau_{\rm exp}$, decay time $t_{\rm D}$, and fit
                time-range, beginning at the time when gas mass starts to decline
                $t_{\rm D}$. The fifth column assess the quality of the fit using
                the ${\cal R}^2$ test, and the last column shows the estimated gas
                expulsion timescale normalized by the cloud initial free fall time.
        \label{tab:tauexp}
}
\begin{tabular}{rc cccc}

\toprule
Model   &  $\tau_{\rm exp}$ & $t_{\rm D}$  & $t_{\rm max}$ &  ${\cal R}^2$ & $\tau_{\rm exp}/t_{\rm ff} $\\
        & [Myr]  & [Myr]  & [Myr] &   &  \\ \hline
\fiducial    & $1.7    \pm  0.1 $ &  2.53  &  9.50   &  0.90 & $0.48   \pm 0.03 $  \\ 
\alphaL      & $1.4    \pm  0.1 $ &  2.78  &  7.50   &  0.90 & $1.05   \pm 0.05 $  \\ 
\alphaH      & $3.9    \pm  0.2 $ &  1.26  & 14.8    &  0.96 & $0.39   \pm 0.03 $  \\ 
\rsmall      & $0.55   \pm  0.02$ &  0.249 &  2.50   &  0.94 & $0.90   \pm 0.04 $  \\ 
\rlarge      & $1.73   \pm  0.09$ & 38.7   & 51.1    &  0.79 & $0.090  \pm 0.005$  \\ 
\Bten        & $2.6    \pm  0.2 $ &  2.27  & 11.5    &  0.91 & $0.72   \pm 0.04 $  \\ 
\Bhundred    & $2.4    \pm  0.5 $ &  3.66  &  7.60   &  0.95 & $0.6    \pm 0.1  $ \\ 
\ISRFten     & $2.4    \pm  0.2 $ &  3.16  & 10.5    &  0.89 & $0.64   \pm 0.05 $ \\ 
\ISRFhundred & $1.6    \pm  0.1 $ &  3.16  &  8.90   &  0.94 & $0.44   \pm 0.03 $ \\ 
\Zten        & $1.43   \pm  0.06$ &  3.66  & 10.8    &  0.49 & $0.38   \pm 0.02 $ \\ 
\Zhundred    & $0.79   \pm  0.06$ &  4.79  &  8.21   &  0.84 & $0.21   \pm 0.02 $  \\
\bottomrule
        \end{tabular}
\end{table}

One important parameter that has remained largely unconstrained in $N$-body studies
is the gas depletion timescale, $\tau_{\rm M}$. Its critical importance comes from
the fact that star cluster formation is inefficient. As star clusters transition
from their embedded stage into a gas-free phase, most of the gravitational energy
binding the cluster together vanishes with the gas. If the transition happens
rapidly, quicker than a crossing time, stars do not have sufficient time to adapt
their orbits to the new potential, and a large fraction of them suddenly find
themselves unbound. Previous studies have left $\tau_{\rm M}$ as a free parameter
or used an arbitrary
value.\footnote{ Usually $\tau_{\rm M}$ is taken as zero, representing
        instantaneous gas expulsion. In this approach the resulting bound fractions
        are interpreted as lower limits.} Here we measure this parameter directly
from the \starforge\ simulations, setting theoretical constraints for gas
expulsion timescales.

As shown in Sec.~\ref{sec:backgroundgas}, we adopt an exponential decay functional
form to model the expulsion of residual gas from the cluster, which is the standard
practice in the $N$-body literature. However, $>90\%$ of the gas is cleared out of
the system by radiation and winds before the first supernovae occurs
\citepalias{Guszejnov2022} and before we begin the $N$-body modeling. We refer 
to this gas removal timescale as dispersal time. Therefore, it
is possible to calculate the dispersal time, which governs most of the gas
dispersal, directly from the {\starforge} calculation, which we do here. 
We refer to this self-consistent dispersal timescale as $\tau_{\rm exp}$, which is
different than the one we use in the $N$-body modeling, $\tau_{M}$, to describe the
removal of the remaining residual gas.

\begin{figure}
        \includegraphics[width=\columnwidth]{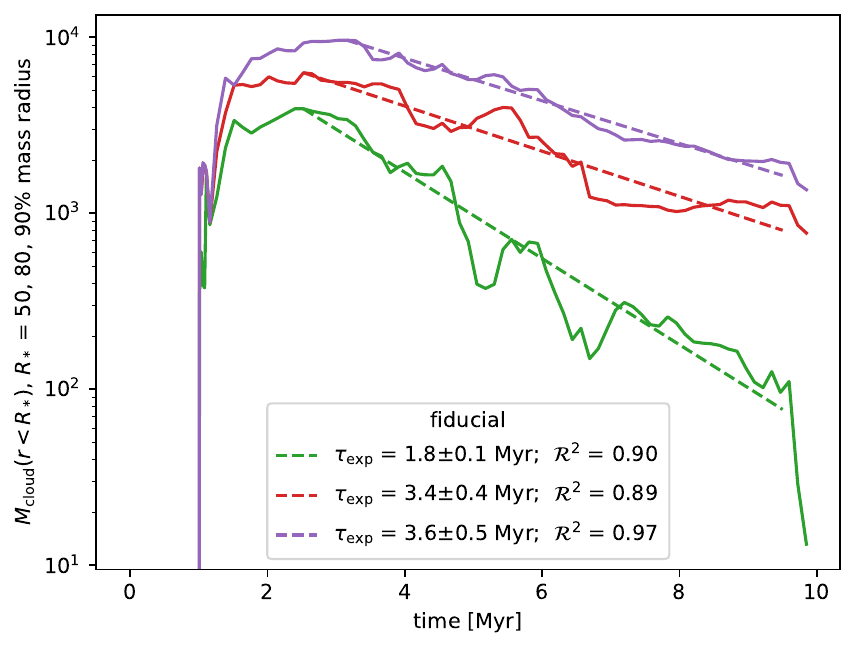} 
        \caption{Gas mass enclosed within the 10, 50, 80 and 90\% stellar mass
        radius during the \fiducial \starforge simulation. Dashed lines shows the
        standard exponential decay timescale fit to the gas evolution.}
        \label{fig:enclosedgas}
\end{figure}

To estimate $\tau_{\rm exp}$, we first measure the gaseous mass enclosed inside the
50, 80 and 90\% radius of the stellar component. Then, we fit the exponential decay
model described by Equation~\ref{eq:gasmass} using a least squares method. In this
procedure, we fit $\tau_{\rm exp}$ as the single free parameter. We select the
scale mass $M_{\rm gas, 0}$ and delay time $t_{\rm D}$ based on the time when the
enclosed gas mass reaches a maximum. Note that we use the gas mass enclosed by the
stars, so the point where the maximum gas is enclosed depends on the forming
stellar distribution and is by definition less than the total cloud mass. We fit
Equation~\ref{eq:gasmass} to the range between $t_{\rm D}$ and the time just before
the explosion of the first supernova, at which point the remaining enclosed gas
mass declines sharply. Figure~\ref{fig:enclosedgas} shows this fitting procedure
applied to the \fiducial\ simulation.

We report the  resulting fit parameters for the gas enclosed within the half mass
radius of the stars in Table~\ref{tab:tauexp} and compare the result to the initial
free fall time of the cloud. We assess the goodness of the fit using the 
${\cal R}^2$ test, which compares the fitting function residuals to the residuals
obtained by using a data average over the fitting range. We find that the
exponential decay model in Equation~\ref{eq:gasmass} is, in general, a good
description of the gas removal that occurs due to radiation and wind feedback. This
is corroborated by the high ${\cal R}^2$ values, considering only one degree of
freedom is given to the fit. However, in some cases the erratic evolution of the
enclosed gas precludes a good fit, such as the case of \Zten, which has ${\cal
R}^2=0.49$. The goodness of fit depends on the location of the star cluster center
with respect to the gas, as well the relative size of both components. Cases such
as the \Zten\ model, where stars expand faster than the gas for a period of time,
or instances where the stars move through overdense regions compromise the goodness
of the description. 

Normalizing by the initial free-fall time of the cloud indicates an order of
magnitude spread in the characteristic timescale. The \alphaH model, where
$\tau_{\rm exp}/t_{\rm ff} = 1.05  \pm 0.05$, has the longest relative
gas-depletion time. This model has a long formation time such that the star cluster
maintains a relatively constant size (see Figure~\ref{fig:revol}). The stellar
distribution in this model takes longer to contract, as the orbits are more
energetic, and it takes more time for the gas to be cleared from a larger volume.
The \rlarge\ model is at the opposite extreme; its depletion timescale is less than
10\% of its free fall time. However, the relatively short depletion timescale
explains why this model forms no bound systems (see Fig.~\ref{fig:grouppar}).
Models with stronger external radiation fields and lower dust abundance, i.e.,
those with warmer gas temperatures and more efficient stellar feedback, also show
short depletion timescales as we would expect.

\begin{figure*}
        \includegraphics[width=\textwidth]{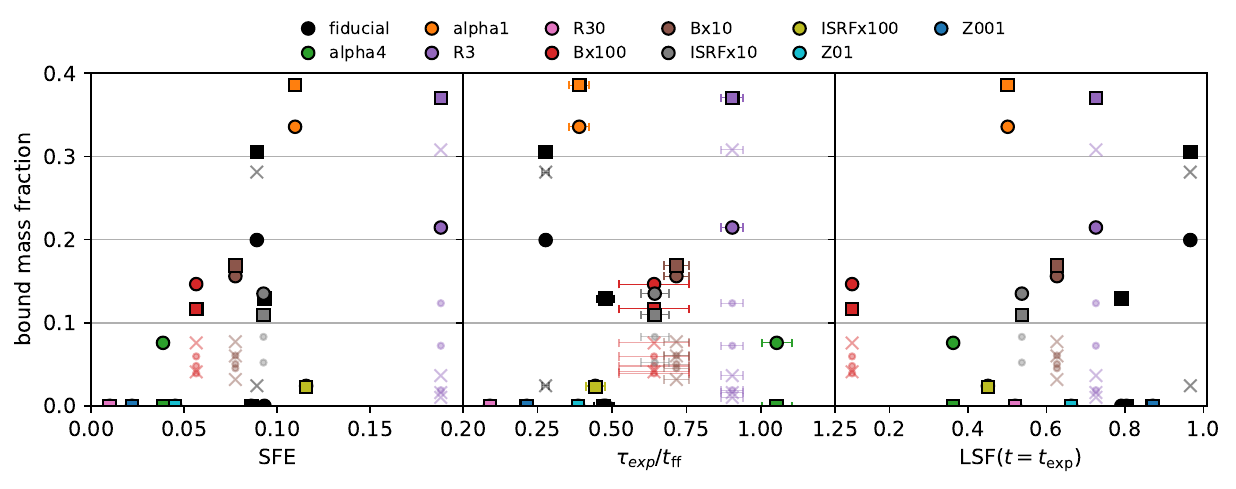}
        \caption{Total bound fraction as a function of global SFE (left), and gas
                expulsion timescale $\tau_{\rm exp}$ normalized by the cloud
                initial freefall time (middle) local, stellar fraction (LSF) at the
                moment of cluster expansion (right). Filled circles and squares
                show global bound fractions for \noextpot\ and \extpot\
                simulations, respectively. Analogously, semitransparent dots and
                crosses show bound fractions for individual clusters.  
        }
        \label{fig:boundfraction}
\end{figure*}
\subsubsection{Bound fractions}\label{sec:boundfraction}

As shown above, the simulated clouds often form more than one bound star cluster and
these are generally a subset of the stars identified in stellar groups. In this
context, we define {\it bound stars}  as stars that are part of any of the
identified clusters within a given simulation. Then, the \emph{total} bound (mass)
fraction refers to the ratio between the total number (mass) of bound stars in a
model relative to all stars formed in that simulation. We use the term
\emph{individual} bound fraction to compare one cluster to the total mass or number
of stars in its parent region.

Figure~\ref{fig:boundfraction} compares the total bound fractions obtained 10 Myr
after the expansion with the global SFE, LSF measured at $\texp$, and explusion
timescale $\tau_{\rm exp}$. We observe a strong correlation between the bound
fraction and the global SFE. Interestingly, within the range of SFE values
obtained from the \starforge\ simulations (i.e.\ SFE$<0.2$), classical models do
not predict the formation of \emph{any} bound systems in the instantaneous gas
expulsion limit. Here, more gradual gas expulsion likely facilitates cluster
formation at lower SFE.
Previous numerical studies have also demonstrated that substructure in the
stellar distribution increases the likelihood of forming bound systems in low SFE
environments \citep[see e.g.,][]{Smith2013,Farias2015}. 
The middle panel of Fig.~\ref{fig:boundfraction}, however, reveals that the bound
fraction increases roughly linearly with normalized $\tau_{\rm exp}$, in agreement
with classical predictions.  However, three models notably deviate from the
relationship: the \alphaL, \alphaH, and one of the \fiducial runs.

The \alphaL\ and \alphaH models suggest that the turbulent velocities and degree of
boundedness of the parent cloud significantly influence the bound fraction. The
\alphaL\ simulation yields the highest bound fraction among all models, despite
having a relatively low SFE (0.07) and rapid gas expulsion compared to the cloud's
free-fall time ($\tau_{\rm exp}/t_{\rm ff} = 0.38$). In contrast, the \alphaH model
has both a low bound fraction and a low SFE, but its expulsion timescale is the
longest among the models. Consequently, its position in
Figure~\ref{fig:boundfraction} lies below the other models. It is conceivable that
these two models follow distinct relations between bound fraction, SFE, LSF, and
$\tau_{\rm exp}/t_{\rm ff}$. The case of the outlier \fiducial\ run, however,
underscores that star cluster formation is a stochastic process and that other
factors may come into play.

Previous numerical studies
\citep[e.g.,][]{Goodwin2009,Smith2011b,Smith2013,Farias2015} have argued that
parameters such as the LSF and the virial parameter at the \emph{moment of gas
expulsion} yield improved estimates of the post-gas-expulsion bound fraction
compared to the global SFE. However, the gas expulsion process, as discussed in
\S\ref{sec:taum}, occurs over a prolonged timescale that begins early in the star
formation process. Hence, the exact moment of gas expulsion is ambiguous.
It's important to note that by the time gas expulsion is considered to ``begin,''
less than half of all stars have actually formed. An alternative reference point
could be the onset of expansion (\texp) of the stellar region. At this point, stars
have undergone significant dynamical relaxation, and it is likely that the majority
of the dynamical interactions that stars will experience have already occurred.

The last panel in Fig.~\ref{fig:boundfraction} shows that  LSF measured at \texp\
also imperfectly correlates with the bound fraction.
In this case, the outlier \fiducial\  in the previous panels has the highest LSF of
all the models at $t=\texp$. However, it is worth noting that the bound fractions
obtained for clusters with high LSFs are significantly lower than those found in
$N$-body experiments, which assume instantaneous gas expulsion \citep[see
e.g.][]{Farias2015,Farias2018b}. 

On the other hand, we observe no correlation between the virial parameter of the
stars and the bound fraction for either gas expulsion definition. This lack of
correlation may be due to the longer timescales over which gas expulsion occurs.
The virial parameter exhibits significant variations during the cluster's formation
(see Figure~\ref{fig:evol_bound}), and there is no single point in time that can
accurately represent the cluster's overall dynamical state.

Altogether, these results imply that the virial state of the parent cloud, along
with the SFE and gas expulsion timescale, plays a crucial role in the formation of
bound clusters.

\subsubsection{Primordial kinematics}

\begin{figure*}
        \includegraphics[width=\textwidth]{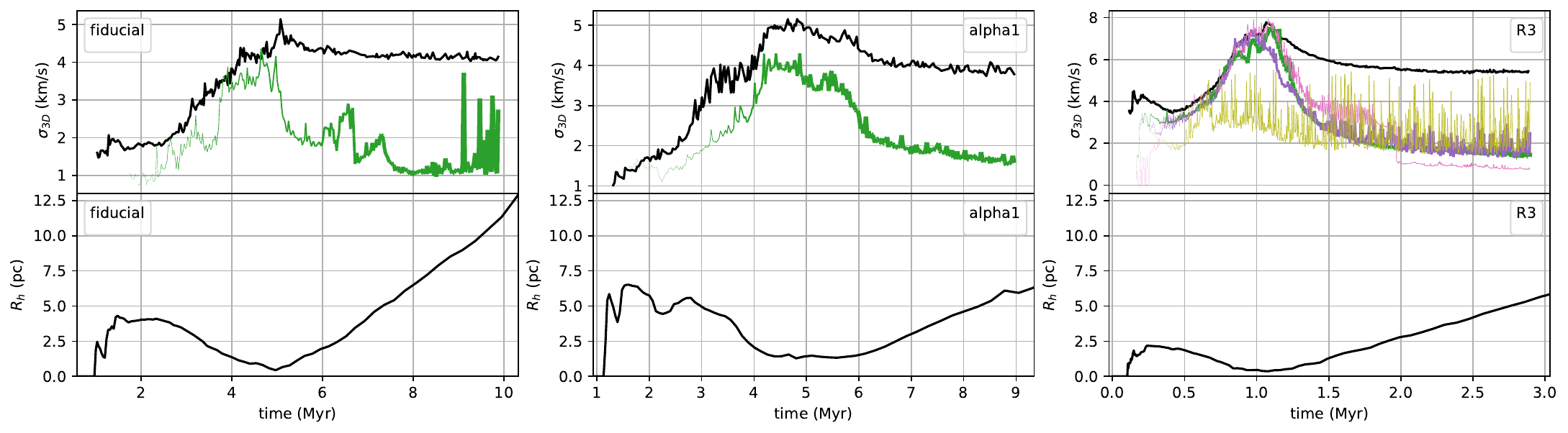}
        \includegraphics[width=\textwidth]{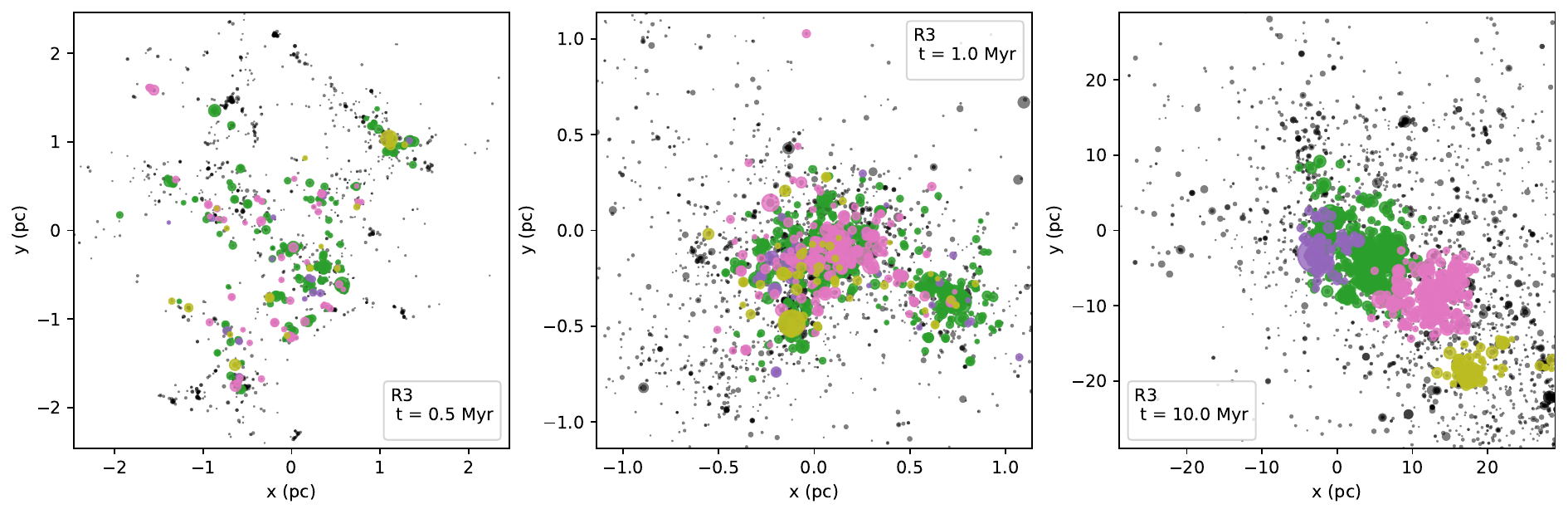}
        \caption{\textbf{Top:} Velocity dispersion for three \starforge\ models as
                a function of time for all stars (black solid lines) and for
                members of bound clusters found at 20 Myr after the expansion of
                the region shown in different colors. To calculate the velocity
                dispersion we use the center of mass velocities of binaries and
                higher-order multiples.
        \textbf{Bottom:} Position of bound members at 0.5, 1 and 10 Myr for the
        \rsmall\ case. Colors in the bottom panels match the groups in the
        top \rsmall\ panel. 
        }
        \label{fig:starforgesigma}
\end{figure*}

Despite a diverse set of initial conditions, we find that the cluster kinematic
properties, particularly the velocity dispersions, are remarkably similar only 10
Myr after the expansion of the stellar complex begins. This raises the question: do
star clusters inherit kinematic properties of the parent cloud or do they form with
independent velocity signatures? To address this question, we trace back bound
members from the identified star clusters and examine their velocity dispersion
histories and places of birth. Figure~\ref{fig:starforgesigma} displays the
evolution of $\sigma_{\rm 3D}$ for the \fiducial, \alphaL, and \rsmall models.

Figure~\ref{fig:starforgesigma} shows that stars destined to form bound clusters
are born with slightly lower velocity dispersions compared to other stars. The
initial velocities of the stars are correlated with the parent cloud velocity
dispersion. For the \fiducial, \alphaL, and \rsmall\ cases shown in the figure, the
initial cloud velocity dispersions are 3.2\,\kms, 2.3\,\kms, and 5.8\,\kms,
respectively \citepalias[see][]{Guszejnov2022}, while the early stellar velocity
dispersion is about a factor of 2 lower. However, the stellar dispersion quickly
increases as the stars become centrally condensed, exceeding the initial cloud
dispersion and reaching a constant value. The kinematics of bound members remain
coupled to the rest of the region until shortly after the complex begins to expand,
although their velocity dispersions are generally smaller than the global stellar
dispersion.

Consequently, we conclude the similar velocity dispersions among the bound systems
after expansion are partly due to a selection effect. During the violent relaxation
that occurs during the region's contraction, the least energetic stars tend to
remain together, effectively erasing any signature of their parent cloud's velocity
dispersion.

Traceback of the individual members of the identified star clusters demonstrates
that the final bound members do not necessarily originate from the same local
region within the cloud. The bottom panels of Figure~\ref{fig:starforgesigma}
illustrate the distribution of bound members in model $\rsmall$ at times close to
their formation (0.5 Myr), when the stars are most centrally condensed (1 Myr), and
at a later time when the bound clusters are fully formed and distinct (10 Myr). At
early times the stars appear well-mixed, and there is no clear signature of their
later cluster assignment. This indicates that bound clusters emerge from the
dynamical relaxation and energy exchange between stars, wherein the least energetic
stars come together in one or more clusters while the most energetic stars expand.
This relaxation process erases any lingering kinematic signature of the parent
cloud. 

\section{Discussion}\label{sec:discussion}
\subsection{Clusters versus associations}

The \starforge\ framework has provided a more self-consistent star cluster
formation model than has previously been available, resolving individual stars and
include all key feedback mechanisms. These processes are crucial for understanding
the early dynamical state of young stellar groups and the process by which star
clusters transition from embedded to gas-free entities. We find that in general the
\starforge simulations form expanding stellar complexes that contain smaller bound
clusters and unbound associations on a range of scales. Given the low global SFE in
the simulations, the total bound mass contained in all modeled stellar complexes is
below 1000\:\msun, $<40\%$ of total stellar mass, with an average of 20\%; this is
either contained in a single cluster or spread between several sub-clusters.
Unbound associations, which may be a mixture of bound and unbound stars, naturally
contain larger mass fractions compared to the total mass or between 20\% and 80\%
all stars in the complex. Given that our models do not account for tidal disruption
from the host galaxy or nearby GMCs,  which would decrease the fraction of
surviving clusters, our mass fractions represent upper limits
\citep{Kruijssen2012a,Kruijssen2012b}.

Our main takeaway is that, while most star systems form with a high degree of {\it
clustering}, these systems primarily form in (unbound) associations not (bound)
clusters. The detailed kinematics available from the \starforge\ simulations
underscore that most stars are {\it not} born into clusters, at least in the
galactic conditions that we have modeled. This conclusion is consistent with recent
{\it Gaia} observations, which have provided a wealth of kinematic data of young
stellar complexes, thereby allowing a clear distinction to be made between clusters
and associations \citep{Chevance2022}. Debate on the topic of cluster
vs.~association has been muddied over the years by different definitions of what
constitutes a star ``cluster," a term that has historically depended on some
predefined separation scale and/or stellar density
\citep[e.g.,][]{Lada2003,Gutermuth2009a}.
The significant differences we find between stellar groups, defined using spatial
information, and clusters, identified using full kinematic information, underscore
the challenges of observationally identifying young clusters and predicting their
evolutionary outcomes without accurate spatial and kinematic information.

\subsection{Implications for massive cluster formation}

We demonstrate in this study that clusters are formed with a range of masses and
densities, and their kinematic properties depend more on their individual masses
than the initial conditions of their parent cloud (see below). By extrapolating
trends shown in Fig.~\ref{fig:alphamass}, we anticipate that denser and/or
more-massive clouds than those considered are needed to produce more massive and
long-lived clusters. This has been found in previous, lower-resolution GMC and
galaxy simulations accounting for stellar feedback
\citep{Li2019,Grudic2021a}, but this regime still warrants exploration with
a more self-consistent numerical treatment of star formation.
       
However, there are also secondary effects that influence the final cluster mass. As
shown in Fig.~\ref{fig:boundfraction} higher SFE indeed correlates with the
formation of higher-mass clusters. SFE in turn correlates with high-surface
densities and lower initial gas velocity dispersions. Some amount of residual gas,
at the few percent level, also boosts the final cluster mass. This suggests there
are multiple  variables that interact to promote the formation of high-mass
clusters. Additional simulations of clouds with realistic feedback processes are
necessary to fully explore the parameter dependencies and investigate the more
extreme conditions likely required to form massive clusters.

\subsection{Group Selection}
In contrast to some recent observational studies
\citep[e.g.][]{Hunt2021,Kerr2021,Quintana2023} we do not use velocity information
to identify groups. We find that using velocities for group identification produces
much noisier groupings, which experience more membership changes over time. For
example, if a group contains binaries and higher order multiples, these members are
often rejected from the selection because of their orbital velocity. 

Observationally, it is advantageous to use velocities for selection as it helps to
eliminate field stars that have large velocity differences with the groups of
interest. In addition, most close binaries in the observed sample are not resolved,
and thus populate a long tail in the observed stellar velocity distribution
\citep[e.g.,][]{DaRioTan2017}.  However, in our case the system velocities are
fairly similar and all our stars naturally form within the same cloud complex. We
also have perfect identification of binaries, allowing us to correct for their
velocities during the group and cluster identification process. Consequently, in
our case the primary velocity differences come from the birth expansion rather than
from different galactic orbits. 

In cases where clustering algorithms employ velocity information, our analysis
suggests that the orbital velocities of unresolved binaries cause the exclusion of
cluster members and modify the appearance of any sub-structure. The absence of
complete kinematic information and the ability to correct for orbital motion,
likely causes weakly bound clusters to be identified as associations, particularly
those with low membership and low  stellar density.

\subsection{Dynamical dependence on group mass}

Our stellar groups span two orders of magnitude in mass, thereby allowing us to
investigate variation in group properties as a function of mass. Notably, we
observe that low mass clusters expand less rapidly during the first 100 Myr of
evolution (see Figure~\ref{fig:evol_bound}), however a considerably fraction of
them dissolve during this time. More massive clusters expand more rapidly
initially, driven by dynamical interactions. However, after about 100 Myr, most of
them are relaxed and reach a relatively steady stellar surface density of
$\sim0.05-0.1$\,stars\,pc$^{-2}$. 

We have shown that a ratio of age to crossing time, i.e., the dynamical age $\Pi$,
of unity is consistent with the separation between bound systems and unbound
groups, as proposed by \citep{Gieles2011}. However, this distinction becomes
noticeable only as stellar groups age. Shortly after the expansion of the
complexes, at 10 Myr, both groups and clusters have dynamical ages below or
close to unity. Low-mass clusters exhibit a rapid increase in dynamical age, while
more massive clusters do so more gradually. By 100 Myr, they barely surpass this
threshold. These results highlight the difficulty to assess the boundness of young
expanding star clusters as, while they are bound, they do expand rapidly at shorter
ages. 

The emergence of these trends can be attributed to the remarkably consistent
velocity dispersion of the identified groups. This produces a mass dependence, as
groups of varying masses experience distinct relaxation patterns within this
confined velocity range. Despite initial variations in the bulk gas velocity
dispersion by a factor of three, the stellar velocity dispersions of groups
identified by HDBSCAN after 10 Myr differ by less than a factor of 2 and show
no correlation with the parent cloud's initial conditions (see
Fig.~\ref{fig:grouppar}). The velocity dispersion of the identified clusters within
the groups displays an even narrower distribution  of $0.2 < \sigma_{\rm 3D} <
0.8\,\kms$. Previous theoretical works have noted the low velocity dispersions of
young groups, arguing that protostars inherit their initial velocities from the
motions of the dense gas \citep{Offner2009a,Li2019}. Indeed, observations of dense
cores find that the dispersion of their bulk velocities, as measured in $^{13}$CO,
is about half that of the overall cloud velocity dispersion \citep{KirkPineda2010}.

However, the stellar velocity dispersion does not remain low for long. The process
of violent dynamical relaxation during the global collapse and merging of different
primordial groups results in the decoupling of stars and gas
kinematics (as illustrated in Fig.~\ref{fig:starforgesigma}). Stars reconfigure
through dynamical interactions, with the most energetic stars leaving the system
and the least energetic ones forming configurations closer to equilibrium.
\cite{Sills2018} also found that the velocity dispersion of embedded substructured
star clusters become similarly smooth in phase-space very quickly regardless choice
of initial virial ratio.

In summary, young stellar systems tend to form near virial equilibrium, even in
cases where their parent clouds possess low SFE or high turbulent velocities. While
these findings shed light on the internal structure of clusters, further
simulations are required to validate and expand these results across a broader
parameter space and model more massive clouds. For now, in the context of these
results, the relevant question reduces to what mass each cluster is able to retain
under the different environments rather than how the environment shapes their
internal structure.

\subsection{Gas depletion timescales}

The significance of gas expulsion as a catalyst for star cluster dissolution has
been under scrutiny in recent years \citep[e.g.][]{Kruijssen2012b}. Due to the
absence of well-defined constraints on gas expulsion timescales, past numerical
investigations have often resorted to treating the gas expulsion/depletion
timescale as an unconstrained parameter
\cite[e.g.][]{Brinkmann2017,Farias2019,Dinnbier2019}, employing uncertain
estimations \citep[e.g.][]{Banerjee2013,Dinnbier2020a}, or adopting the assumption
of instantaneous gas expulsion \citep[e.g.][]{Farias2017,Shukirgaliyev2023}.
Consequently, the need for more precise constraints to assess the impact of this
process \citep{Dinnbier2022} becomes evident. 

This work provides self-consistent estimates of the gas depletion timescale across a range
of cloud environments. According to our definition, the gas fraction within
embedded star clusters decreases right from the onset of the star formation
process, influenced by both stellar feedback and the motion of stars as
they concentrate. All the models we examine have gas depletion timescales either
below or comparable to a single free fall time. The exponential decay model states
that within a timescale of one $\texp$, stars deplete 60\% of the gas within their
half-mass radius. On average, this timescale corresponds to half a free-fall time,
very close to the \fiducial\ case, with a maximum of one free-fall time in the
\alphaH case (refer to Table \ref{tab:tauexp}).

The expected correlations would suggest that models with more energetic or
efficient radiative feedback, such as those with lower metallicity and
warmer gas due to higher external irradiation, would have
rapid gas removal than in the fiducial cases \citep[see e.g.][]{Li2019}. However,
these anticipated trends do not manifest prominently in the results, as the
effective gas expulsion timescale, by our definition, hinges on the interplay
between the distributions and motions of stars relative to the gas. Additionally,
the positions of the most massive stars within the complex may exert a
substantial influence on this timescale \citep{Dinnbier2020}. 
Some variation is also expected among different models sharing the same initial
conditions. For instance, within our \fiducial model set of three
simulations, $\tau_{\rm exp}$ varies from 0.7 to 1.5 Myr. 

The normalized gas expulsion timescales we compute align well with the
bound fractions of each region. While some scatter within these relationships could
be attributed to other contributing factors, the dynamical state of the parent
cloud emerges as a pertinent variable. This is evident as distinctive models with
higher and lower virial parameters fall above and below the general
correlation. Another outlier in the correlation, observed in one of the fiducial
models displaying a high bound fraction despite a relatively swift gas expulsion
timescale, can be attributed to the high local stellar fraction reached during the
collapse phase \citep[see e.g.][]{Smith2013}.

\section{Conclusions}\label{sec:conclusion}
In this paper we explored the long-term evolution of stars formed under the
\starforge framework. We evolved the stellar distributions for 200 Myr after
formation using Nbody6++. We used the clustering algorithm HDBSCAN to identify
groups and employed a dynamical analysis to identify true (bound) clusters. We then
derived the properties of the resulting star clusters and associations.  
We summarize our results as follows:
\begin{itemize}
\item  The \starforge\ clouds primarily form unbound expanding complexes that
        include one or more star clusters and (unbound) associations in each
        simulation.

\item Most of stellar groups formed in these models are associations with surface
        densities below 1\:\Sigmaunit. We find the densities of the identified
        groups remain above $10^{-2}\:\Sigmaunit$ for $\sim$100 Myr. 

\item One to four star clusters form in each simulated cloud, with the majority
        having fewer than 200 members. No clusters form in the low metallicity
        clouds (0.1$Z_\odot$, 0.001$Z_\odot$) or the lowest surface density cloud.
        Two clusters are relatively large, with  $\sim$800 and 1100 members.
        These form in clouds with roughly virialized turbulent gas velocities and
        an initial gas density ten times higher than the fiducial model. Notably, a
        comparable large cluster originated in one of the fiducial models that
        experienced a particularly high local stellar fraction during the collapse
        phase.

\item  We find that including a background potential to represent residual gas has a
        small positive impact on the size of identified clusters. In a couple
        cases, the background potential promotes merging,  doubling the size of the
        final cluster. The non-linear dynamics of the stellar interactions means
        that even a small amount of remaining gas, i.e., $<10$\% by mass, can
        produce substantial changes in cluster formation. 

\item  Star clusters form with velocity dispersions between 0.2 and 0.8 \kms,
        tightly correlated with cluster mass. This correlation seems independent of
        parent cloud conditions. Smaller velocity dispersion leads to smaller
        half-mass radii in smaller clusters. Initially, all star clusters expand, with
        larger ones expanding faster until stabilizing around 100 Myr. Lower mass
        clusters expand more slowly, however, 40\% of them dissolve before 100 Myr.

\item The identified stellar groups exhibit a broad range of velocity dispersions,
        spanning from 0.2 to 100~\kms, and do not show any clear correlations with
        mass. In contrast to clusters, their properties and scaling relations
        exhibit large scatter, with little variation over time. This makes it
        difficult to distinguish them from clusters, particularly in early stages;
        however, differences arise as clusters age.

\item We characterized each cloud's gas depletion evolution using an exponential
        decay model. The characteristic gas expulsion timescales in all models are
        below one initial free fall time of the parent cloud with an average of
        half a free-fall time. In general, more efficient stellar feedback reduces
        this timescale, although it's dependence on the relative motions between gas
        and stars can be considerable.

\item The global bound fractions correlate well with
        both SFE and the gas expulsion timescale. Consequently, given that the SFE
        is below 20\% in all models, the bound fractions are also low, below
        40\%.

\end{itemize}

Given that the \starforge\ simulations account for all the key physical processes
involved in star cluster formation while simultaneously resolving the
stellar IMF, we conclude that these simulations describe a viable
formation scenario for the formation of stellar associations and small clusters.
The formation of large clusters requires initially higher mass clouds, denser
clouds or both. 

\section*{Acknowledgments}

JPF and SSRO are supported by NASA grant 80NSSC20K0507  and NSF Career award
1748571. SSRO also acknowledges support from NSF AAG 2107942, NSF AAG 2107340, NASA
grant 80NSSC23K0476 and a Moncrief Grand Challenge Award from the Oden institute.
We thank ChatGPT for its meticulous identification of grammar mistakes.

\section*{Data availability}
The data underlying this article will be shared on reasonable request to the corresponding author.

\bibliographystyle{mnras}
\bibliography{references}
\bsp	%
\label{lastpage}
\end{document}